\documentclass{aa}  
\usepackage{natbib}
\usepackage{amsmath}
\usepackage{xspace}
\usepackage{txfonts}
\usepackage{color}
\usepackage{url}
\usepackage[flushleft]{threeparttable}
\usepackage[colorlinks=true, allcolors=blue]{hyperref}
\usepackage{orcidlink}
\usepackage{adjustbox}
\usepackage{multicol}
\usepackage{multirow}
\usepackage{xspace}
\usepackage{lastpage}

\newcommand {\ixpe}{{IXPE}\xspace}
\newcommand {\nustar}{{NuSTAR}\xspace}
\newcommand {\nicer}{{NICER}\xspace} 
\usepackage{graphicx}
\usepackage{txfonts}
\begin{document} 

   \titlerunning{Phase-Resolved X-ray Polarimetry of LMC~X-1: Enhanced Polarization at Inferior Conjunction}
   
   \title{Phase-Resolved X-ray Polarimetry of LMC~X-1: Enhanced Polarization at Inferior Conjunction}


   \author{Yu-Shan Ling
          \inst{\ref{inst:gx}}\orcidlink{0009-0002-4994-2528}
          \and
          Fei Xie
          \inst{\ref{inst:gx}}\fnmsep\thanks{Corresponding author: xief@gxu.edu.cn}\orcidlink{0000-0002-0105-5826}
          \and
          Alessandro Di Marco
          \inst{\ref{inst:iaps}}\orcidlink{0000-0003-0331-3259}
          \and
          Fabio La Monaca
          \inst{\ref{inst:UTU},\ref{inst:iaps}}\orcidlink{0000-0001-8916-4156}
          }

   \institute{Guangxi Key Laboratory for Relativistic Astrophysics, School of Physical Science and Technology, Guangxi University\label{inst:gx}
         \and
             INAF Istituto di Astrofisica e Planetologia Spaziali, Via del Fosso del Cavaliere 100, I-00133 Roma, Italy\label{inst:iaps}
         \and
         Department of Physics and Astronomy, FI-20014 University of Turku, Finland\label{inst:UTU}
             }

   \date{Received ; accepted }

 
  \abstract
   {The extragalactic black hole X-ray binary LMC~X-1 is a classical high/soft state source with a well-characterized accretion disk and an observed ionization cone/infered jet structure. Previous X-ray polarimetric observations of LMC~X-1 yielded only orbit-averaged upper limits on the polarization fraction, so that any orbital-phase dependence remained unexplored.}
   {We aim to perform a phase-resolved X-ray polarimetric analysis of LMC~X-1 to probe the orbital-phase dependence of polarization and identify the origin of the polarized emission.}
   {We analyzed simultaneous observations from \ixpe, \nicer, and \nustar taken in October 2022. We measured the polarization degree and angle in different energy bands as a function of orbital phase, and performed spectro‑polarimetric modeling to identify the emitting components. We also modeled the phase modulation of the Stokes parameters with a cosine function.}
   {The source exhibits an orbital-phase dependence of polarization: during inferior conjunction ($\phi = [-0.1, 0.1]$), the polarization degree is $(4.2\pm1.1)\%$ in the 2--3\,keV band and $(3.4\pm0.8)\%$ in the 2--8\,keV band, with polarization angles of $(46\pm8)^{\circ}$ and $(50\pm7)^{\circ}$, respectively. Outside this phase, the polarization remains below the 99\% minimum detectable level. The measured polarization angle is consistent with the ionization cone/jet direction (${\sim}45^{\circ}$). Spectro-polarimetric modeling favors a soft band origin for the polarization. The phase modulation of Stokes $Q/I$ and $U/I$ is well described by a cosine at the orbital period, and it tends to require two spatially separated polarized components rather than a single one varying in amplitude. We also tested whether differential Faraday rotation of the disk light in an ordered field can reproduce the modulation and used it to place upper limits on the magnetic field. These results reveal that the soft-band polarization of LMC~X-1 is phase-dependent, appearing only near inferior conjunction and at a position angle aligned with the disk normal, and they highlight the importance of orbital-phase sampling in polarimetric studies.}
   {}

   \keywords{X-ray binary stars --
                Stellar mass black holes --
                Stellar accretion disks --
                Polarimetry
               }

   \maketitle
%
\nolinenumbers 
\section{Introduction}

LMC~X-1, located in the Large Magellanic Cloud, is the first extragalactic black hole X-ray binary ever discovered \citep{Mark1969,Price1971}. It is at a distance of approximately 50\,kpc \citep{Pietrzynski2013}, and its optical counterpart is an O7/8~III giant \citep{Hutchings1983,Negueruela2002}. An orbital period of $3.90917(5)$\,days from high‑resolution optical spectroscopy and an X-ray period of P = $3.9094(8)$\,days from {\it Rossi X-ray Timing Explorer All-Sky Monitor} (RXTE/ASM) data were both determined by \citealt{Orosz2009}, and they are mutually consistent. The black hole (BH) mass is $(10.9\pm1.4)\,M_{\odot}$, with a companion star mass of $(32\pm3)\,M_{\odot}$ and an orbital inclination of $(36.4\pm1.9)^{\circ}$ \citep{Orosz2009}.

The black hole spin has been extensively studied using the continuum‑fitting method, yielding remarkably high values: $a_{*} = 0.92^{+0.05}_{-0.07}$ \citep{Gou2009}, $0.82$--$0.92$ \citep{Bhuvana2021}, $0.92$--$0.95$ \citep{Jana2021}, and the study of Fe‑line fitting the reflection component also gives consistent high spins \citep{Steiner2012}. However, \citet{Zdziarski2024} and \citet{Zdziarski2026} have recently shown that the inferred spin is strongly model-dependent; in fact, introducing a warm, optically thick surface layer reduces the best‑fit spin to $a_{*} = 0.4^{+0.4}_{-0.3}$, highlighting the impact on spin determination of the assumed disk structure.

BHs are usually observed in different spectral states; see, e.g., \citet{Done2007, Marcel2026}. LMC~X‑1 is persistently observed in the high/soft spectral state, making it rare among dynamically confirmed BH candidates \citep{Nowak2001}. Its X‑ray spectrum is dominated by a multi‑temperature disk blackbody component, contributing ${>}80\%$ to the flux, with a weak steep power‑law component ($\Gamma{\gtrsim}2$) from Comptonization \citep{Nowak2001, Gou2009, Hanke2010, Jana2021, Bhuvana2021}.
A broad Fe K$\alpha$ emission line at ${\sim}6.4$\,keV is detected, originating from the relativistic accretion disk \citep{Steiner2012,Jana2021,Bhuvana2021}. The X-ray luminosity is $L_{\mathrm{X}} \approx (1.4$--$1.5) \times 10^{38}\,\mathrm{erg\,s^{-1}}$, corresponding to $L/L_{\mathrm{Edd}} {\sim} 0.07$--$0.24$ \citep{Jana2021,Bhuvana2021}, with an inferred accretion efficiency $\eta{\sim}0.13$ \citep{Jana2021}.
\citealt{Hanke2010} derived equivalent hydrogen column densities of $N_{\mathrm{H}} = (1.0$--$1.3) \times 10^{22}\,\mathrm{cm^{-2}}$ toward LMC~X-1 after accounting for the lower metallicity of the Large Magellanic Cloud. They also found an orbital-phase modulation for $N_{\mathrm{H}}$, with higher values corresponding to the phase in which the BH is behind the donor star, which could explain the observed ${\sim}7\%$ orbital modulation of the soft X-ray flux detected with RXTE/ASM \citep{Orosz2009,Hanke2010}. This orbital modulation of the X-ray flux is well established: RXTE/ASM data over 12 years show a ${\sim}7\%$ sinusoidal variation, fully consistent with the optical period, and attributed to Thomson scattering and absorption in the wind of the companion \citep{Orosz2009,Hanke2010}. This modulation is also reported to be evident in the Neutron Star Interior Composition Explorer (\nicer) and the Imaging X-ray Polarimetry Explorer (\ixpe) light curves \citep{Podgorny2023}.
The source exhibits an ionization cone with a projected position angle of ${\sim}225^\circ$ (or ${\sim}45^\circ$, as the direction is degenerate by $180^\circ$), aligned with the inferred jet direction \citep{Cooke2007, Cooke2008}. The He~II emission requires an ultraviolet--soft X-ray excess from the accretion disk, with the ionizing photon flux exceeding standard X-ray model predictions by a factor of 7--10 \citep{Cooke2008}.

The first X-ray polarimetric observation of LMC~X-1, performed by \ixpe \citep{Weisskopf2022, Soffitta2021} in October 2022, found a polarization degree (PD) of $(1.0\pm 0.4)\%$ in the 2--8\,keV band \citep{Podgorny2023}, below the minimum detectable polarization at 99\% confidence level (MDP$_{99}$) of 1.1\%. This low polarization is consistent with Chandrasekhar's predictions for pure thermal emission from a geometrically thin, optically thick accretion disk seen at low inclination \citep{Chandrasekhar60, Sobolev63, Loktev22}. Moreover, \citet{Podgorny2023} reported hints for orbital-phase-dependent polarization, suggesting that the polarization may vary on timescales comparable to the orbital period. Motivated by this possibility, we performed a deeper time-dependent polarimetric analysis and a joint spectropolarimetric analysis of LMC~X-1 using the simultaneous \ixpe, \nicer \citep{Arzoumanian2014,Gendreau2016}, and Nuclear Spectroscopic Telescope Array \citep[\nustar;][]{Harrison2013} data. In this study, we characterize the short-timescale polarimetric variability and its connection to the orbital and accretion geometry. The paper is structured as follows: Section~\ref{sec:Observation and data reduction} describes the observations and data reduction; Section~\ref{sec:Results} presents the timing, spectral, and polarimetric results, which are discussed with their physical implications in Section~\ref{sec:Discussion}; the main findings and implications are summarized in Section~\ref{sec:conclusions}.

\section{Observation and data reduction} \label{sec:Observation and data reduction}

\ixpe observed LMC~X-1 from 2022-10-19 to 2022-10-28 (observation ID 02001901) for a total effective exposure time of approximately 562\,ks for each of the three Detector Units (DUs). The Level 2 data products were obtained from the HEASARC archive and processed for background rejection \citep{Di_Marco2023}. Then, source events were extracted from a circular region of $90^{\prime\prime}$ radius centered on the source position using SAOImage~DS9 version 7.5 \citep{Joye2003}. The background was estimated from an annular region with inner and outer radii of $150^{\prime\prime}$ and $240^{\prime\prime}$, adopting the same regions as in \citealt{Podgorny2023}.
For the polarimetric analysis, we used the unweighted \texttt{PCUBE} algorithm in the \textsc{IXPEOBSSIM} package version 33.2.1 \citep{Baldini2022}, with the appropriate response matrix (semester 20220702, v013). \ixpe spectra were extracted using the weighted approach \citep{DiMarco22a}, binned to achieve a minimum of 30 counts per energy bin for the Stokes $I$ spectra, while the $Q$ and $U$ spectra have a constant binning of 200\,eV as suggested by \citealt{DiMarco26}. All \ixpe data were analyzed in the 2--8\,keV energy band in the \textsc{xspec} \citep[version 12.15.1;][]{Arnaud1996}, utilizing the \texttt{wilm} abundance \citep{Wilms2000} and the \texttt{vern} cross sections \citep{Vern95}.

\nicer simultaneously observed LMC~X-1 during the \ixpe campaign across 10 observations (ObsIDs from 5100070101 to 5100070110) between the period from October 19 to October 28, 2022. The total exposure time was approximately 25\,ks. The data were processed with the \texttt{nicerl2} and \texttt{nicerl3-spect} standard pipeline in HEASoft version 6.36. Good Time Intervals (GTIs) were obtained by discarding periods with abnormally high count rates in the 12--15\,keV band. The \nicer spectra were extracted in the 1--10\,keV band and optimally binned following the prescription of \citet{Kaastra2016}, requiring a minimum of 30 counts per bin.

\nustar conducted a single observation of LMC~X-1 on 24 October 2022 (ObsID 90801324002), with a total exposure of 19\,ks. Source events were extracted from circular regions of $80^{\prime\prime}$ radius centered on the source position using SAOImage DS9 \citep{Joye2003}, for both FPMA and FPMB. Background spectra were extracted from regions of the same size located away from the source. The spectra were extracted using the standard \texttt{nuproducts} pipeline. The \nustar spectra were fitted in the 3--40\,keV energy range to avoid energies with dominating background contribution and also binned following the prescription of \citet{Kaastra2016}, requiring a minimum of 30 counts per bin.

\section{Results} \label{sec:Results}

\begin{figure}[!b]
\centering
\includegraphics[width=0.8\linewidth]{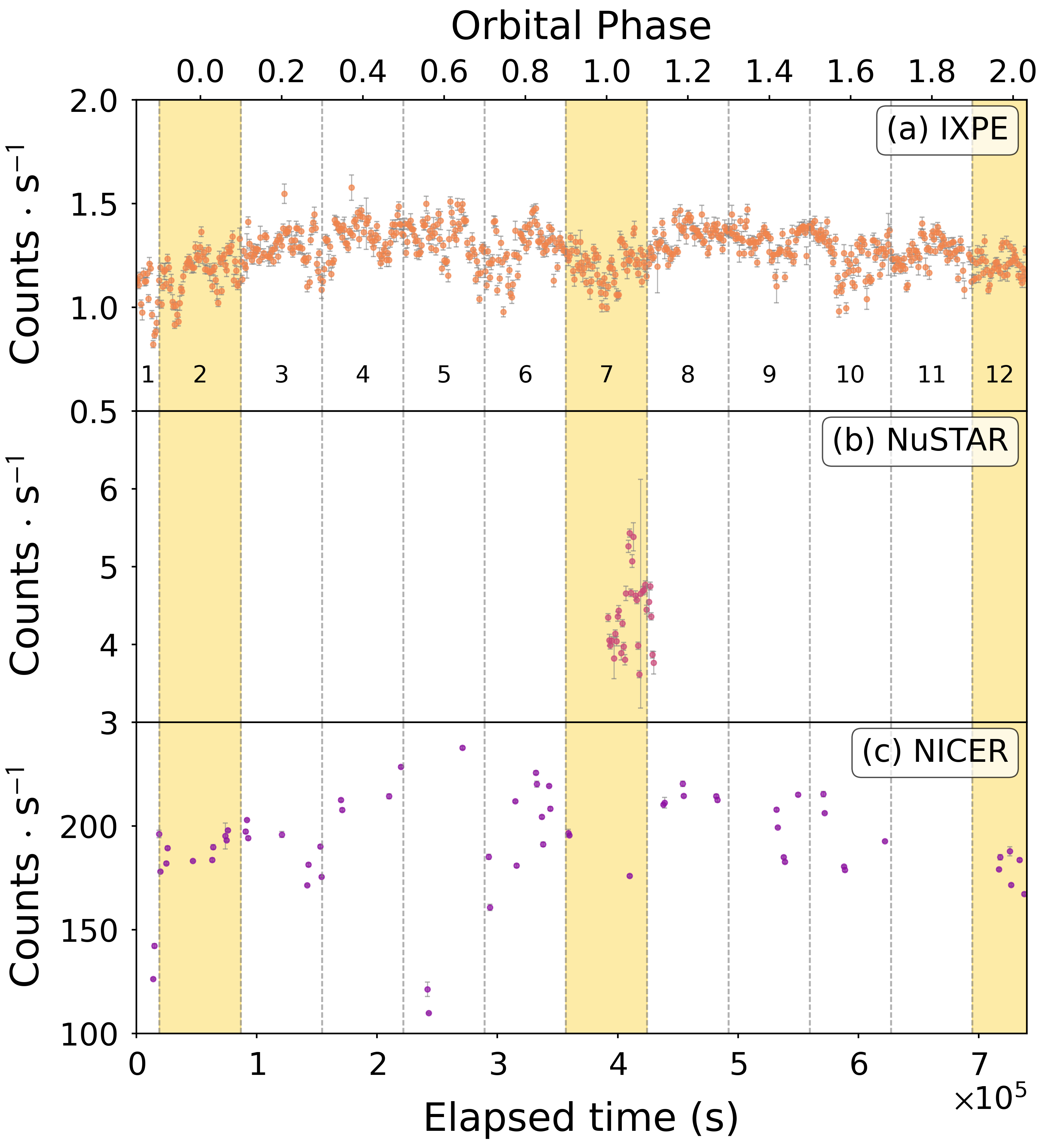}
\caption{X-ray light curves of LMC~X-1 from \ixpe, \nicer, and \nustar, obtained during the 2022 October observations, with a bin size of 1000\,s. The yellow shaded regions mark the interval we chose to identify the inferior conjunction ($\phi = [-0.1,0.1]$), while the gray dashed lines divide the observations into 0.2 phase intervals.}
\label{fig:light_curves}
\end{figure}

To investigate the orbital-phase dependence of the X-ray polarization, we folded the \ixpe events using the orbital ephemeris from \citet{Orosz2009}, with a period of $\textrm{P} = 3.90917(5)$\,days and a reference time of inferior conjunction at MJD~53390.8436. As reported in Figure~\ref{fig:light_curves}, the \ixpe observation covered more than two complete orbital cycles, providing adequate exposure for phase-resolved polarimetric analysis. 
The \ixpe and \nicer light curves show the presence of an orbital modulation, as shown in previous X-ray monitoring, with a minimum occurring around the inferior conjunction of the black hole at phase $\phi = 0$ \citep{Orosz2009}. \nustar, on the other hand, covered only the orbital phase near the inferior conjunction, which means that we have no broadband spectral coverage at other orbital phases.

\begin{figure}
\centering
\includegraphics[width=0.8\linewidth]{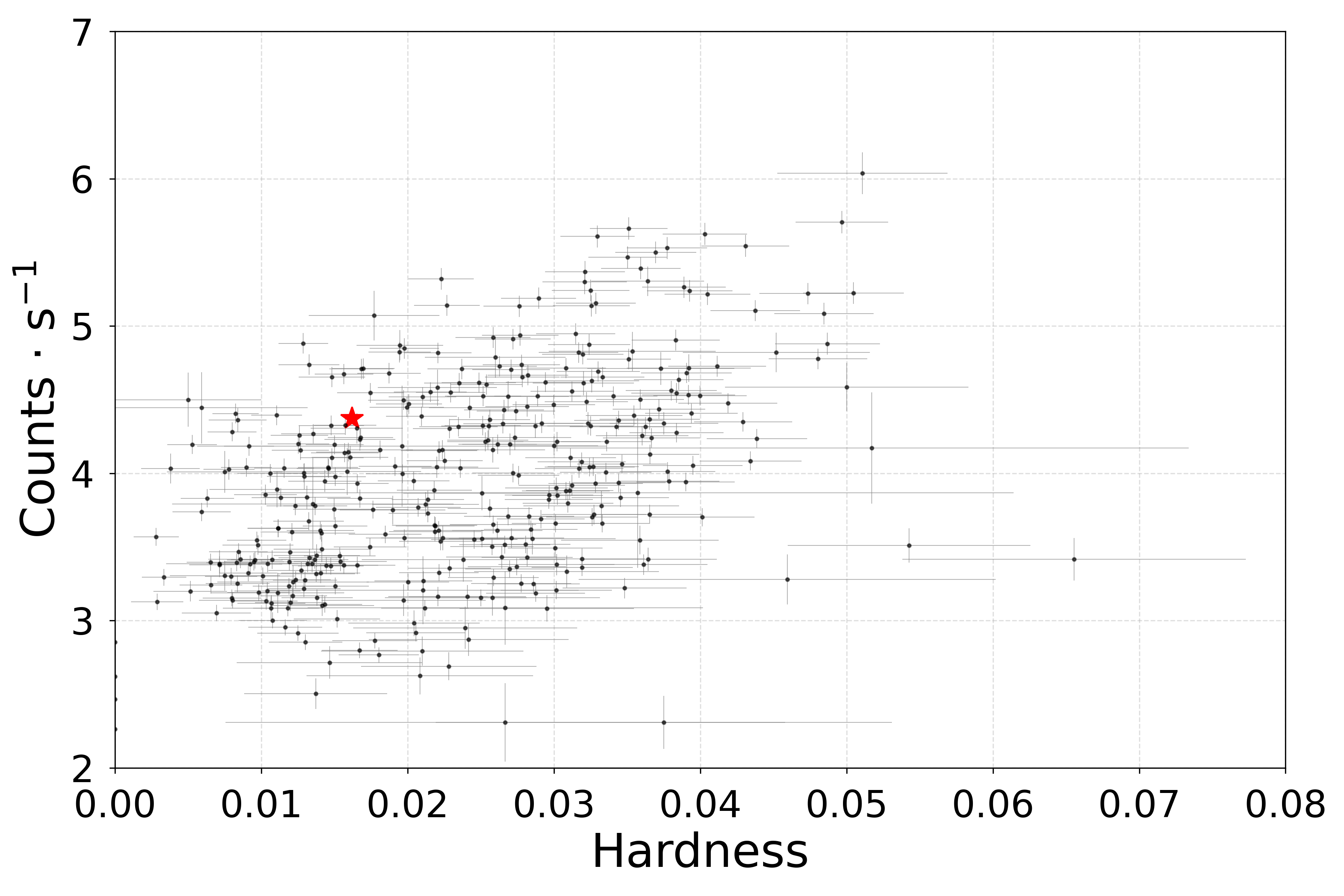}
\caption{HID for LMC~X‑1 as obtained by all the NuSTAR observations available on HEASARC. The hardness is defined as the ratio of counts in the 12--40\,keV to 3--12\,keV bands with a bin size of 500\,s. The red star highlights the observation 90801324002 performed in 2022.
}
\label{fig:HID}
\end{figure}

Figure~\ref{fig:HID} displays the hardness–intensity diagram (HID) of LMC~X‑1, based on all 6 \nustar observations from 2014 to 2022. Each gray point corresponds to one time bin of 500\,s, and the red star marks the time‑averaged state of the 2022 observation (obsID 90801324002). The HID shows that LMC~X‑1 remains in a persistently high‑soft state across all observations, with little change in both hardness and intensity. This spectral stability is consistent with the well‑known behavior of this source, which rarely exhibits significant state transitions \citep{Nowak2001}.

Considering that the inferior conjunction corresponds to a phase with orbital geometry causing the X-ray source to pass behind denser regions of the O-star's stellar wind, such a geometry may induce polarimetric variations, as proposed by \citet{DiMarco25} and \citet{DiMarco26}, including a rotation of the PA and an enhancement of the PD. Due to this specific variation in the X-ray emission visible along the line of sight, we divided the orbital cycle into different time/orbital intervals and into two main intervals: the inferior conjunction ($\phi = [-0.1,0.1]$, where the flux is lowest) and the non-inferior conjunction ($\phi = [0.1,0.9]$), covering the remaining orbit phases (see Figure~\ref{fig:light_curves}). The polarization properties were then studied across the entire orbital cycle and also compared between the two main intervals to search for orbital phase-dependent polarimetric variability.

\subsection{Model independent polarimetric results} \label{subsec:Model independent polarimetric results}

Firstly, we applied the \texttt{PCUBE} algorithm, available in \textsc{ixpeobssim}, to the entire observation to check the consistency of our results with the previous study reported in \citet{Podgorny2023}. The measured polarization is $(0.8\pm0.4)\%$ in 2--8\,keV (hereafter uncertainties are reported at 68\% confidence level -- CL -- if not otherwise stated), below the minimum detectable polarization at 99\% (MDP$_{99}$) of 1.2\%. The result is in agreement with the previous study, although new software and response functions are applied.

\begin{figure}[!ht]
\centering
\includegraphics[width=0.8\linewidth]{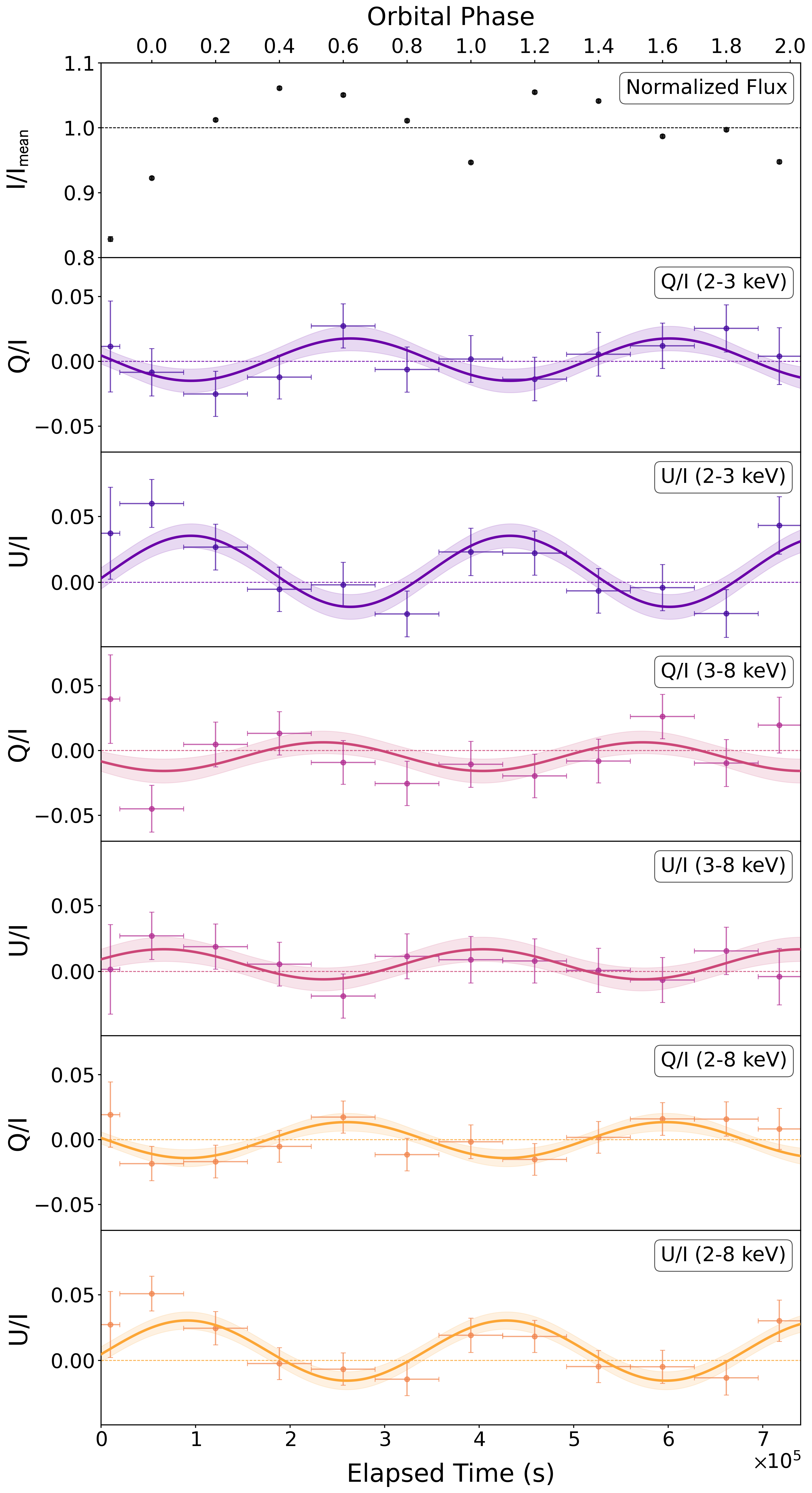}
\caption{From top to bottom: normalized flux in the 2--8\,keV, $Q/I$ and $U/I$ versus time in the 2--3\,keV, 3--8\,keV, and 2--8\,keV energy bands. Data are extracted from time bins covering 0.2 orbital phases in width (except for the first and last bins); errors are reported at 68\%\,CL. The solid-line curves are the best-fits with shaded regions reporting 68\%\,CL allowed bands.}
\label{fig:QU_vs_time}
\end{figure}

After verifying consistency, we investigated the orbital modulation of the polarization in time bins covering 0.2 orbital phases (except for the first and last bins) and for different energy bands; Figure~\ref{fig:QU_vs_time} and Table~\ref{tab:QU_Parameters} report the normalized flux in the 2--8\,keV, and the Stokes parameters $Q/I$ and $U/I$ in the 2--3\,keV, 3--8\,keV, and 2--8\, keV energy bands. Trying to model the behavior of this time dependency, a cosine model, similar to the one applied by \citet{Kravtsov2025}, has been applied, with the period ($P$) fixed to the orbital value, the phase offset fixed to $\phi_0$; the polarimetric amplitudes $A_{Q/U}$, and constants $C_{Q/U}$, were left free to vary.
\begin{equation}\label{eq:QUcos}
\begin{aligned}
(Q/I)(t) &= A_Q \cos\left(2\pi(\frac{t}{P} + \phi_0)\right) + C_Q\\
(U/I)(t) &= A_U \cos\left(2\pi(\frac{t}{P} + \phi_0) -\pi\right) + C_U
\end{aligned}
\end{equation}
The best-fit parameters are reported in Table~\ref{tab:cos_Parameters}. We also performed the same analysis on phase-folded data, and the results are consistent (see Appendix~\ref{app:Folded_QU} for details).
\begin{table*}
\centering
\caption{$Q/I$ and $U/I$ of every time interval in the 2--3\,keV, 3--8\,keV, and 2--8\,keV bands. The definition of time bin refers to Figure~\ref{fig:light_curves}. Errors are reported at 68\%\,CL.}
\label{tab:QU_Parameters}
\begin{tabular}{l c c c c c c}
\hline
\multicolumn{1}{c}{Time bin} & \multicolumn{2}{c}{2--3\,keV} & \multicolumn{2}{c}{3--8\,keV} & \multicolumn{2}{c}{2--8\,keV} \\
\cline{2-7}
 & $Q/I$ (\%) & $U/I$ (\%) & $Q/I$ (\%) & $U/I$ (\%) & $Q/I$ (\%) & $U/I$ (\%) \\
\hline
1 & $1.1\pm3.5$ & $3.7\pm3.5$ & $3.9\pm3.4$ & $0.2\pm3.4$ & $1.9\pm2.5$ & $2.7\pm2.5$ \\
2 & $-0.9\pm1.8$ & $6.0\pm1.8$ & $-4.5\pm1.8$ & $2.7\pm1.8$ & $-1.9\pm1.3$ & $5.1\pm1.3$ \\
3 & $-2.5\pm1.7$ & $2.7\pm1.7$ & $0.5\pm1.7$ & $1.9\pm1.7$ & $-1.7\pm1.3$ & $2.4\pm1.3$ \\
4 & $-1.2\pm1.7$ & $-0.5\pm1.7$ & $1.3\pm1.7$ & $0.5\pm1.7$ & $-0.5\pm1.2$ & $-0.2\pm1.2$ \\
5 & $2.7\pm1.7$ & $-0.2\pm1.7$ & $-0.9\pm1.7$ & $-1.9\pm1.7$ & $1.7\pm1.2$ & $-0.7\pm1.2$ \\
6 & $-0.6\pm1.7$ & $-2.4\pm1.7$ & $-2.6\pm1.7$ & $1.1\pm1.7$ & $-1.2\pm1.3$ & $-1.4\pm1.3$ \\
7 & $0.2\pm1.8$ & $2.3\pm1.8$ & $-1.1\pm1.8$ & $0.9\pm1.8$ & $-0.2\pm1.3$ & $1.9\pm1.3$ \\
8 & $-1.4\pm1.7$ & $2.2\pm1.7$ & $-2.0\pm1.7$ & $0.8\pm1.7$ & $-1.5\pm1.2$ & $1.8\pm1.2$ \\
9 & $0.5\pm1.7$ & $-0.7\pm1.7$ & $-0.8\pm1.7$ & $0.1\pm1.7$ & $0.2\pm1.2$ & $-0.5\pm1.2$ \\
10 & $1.2\pm1.8$ & $-0.4\pm1.8$ & $2.6\pm1.7$ & $-0.7\pm1.7$ & $1.6\pm1.3$ & $-0.5\pm1.3$ \\
11 & $2.5\pm1.8$ & $-2.4\pm1.8$ & $-1.0\pm1.8$ & $1.6\pm1.8$ & $1.6\pm1.3$ & $-1.3\pm1.3$ \\
12 & $0.4\pm2.2$ & $4.3\pm2.2$ & $1.9\pm2.1$ & $-0.4\pm2.1$ & $0.8\pm1.6$ & $3.0\pm1.6$ \\
\hline
\end{tabular}
\end{table*}
\begin{table}
\centering
\caption{Best-fit parameters of the cosine model for the $Q/I$ and $U/I$ Stokes parameters in the 2--3\,keV, 3--8\,keV, and 2--8\,keV bands. Errors are reported at 68\%\,CL.}
\label{tab:cos_Parameters}
\begin{tabular}{lcc}
\hline
Energy & Parameter & Value \\
\hline
2--3\,keV & $A_{Q}$ & $0.016\pm0.008$ \\
 & $A_{U}$ & $0.027\pm0.008$ \\
 & $\phi_0$ & $0.22\pm0.04$ \\
 & $C_{Q}$ & $0.001\pm0.005$ \\
 & $C_{U}$ & $0.008\pm0.005$ \\
 & $\chi^{2}/d.o.f.$ & 14/19 \\
\hline
3--8\,keV & $A_{Q}$ & $0.011\pm0.007$ \\
 & $A_{U}$ & $0.011\pm0.007$ \\
 & $\phi_0$ & $0.30\pm0.07$ \\
 & $C_{Q}$ & $-0.005\pm0.005$ \\
 & $C_{U}$ & $0.005\pm0.005$ \\
 & $\chi^{2}/d.o.f.$ & 16/19 \\
\hline
2--8\,keV & $A_{Q}$ & $0.014\pm0.006$ \\
 & $A_{U}$ & $0.023\pm0.006$ \\
 & $\phi_0$ & $0.23\pm0.03$ \\
 & $C_{Q}$ & $-0.001\pm0.004$ \\
 & $C_{U}$ & $0.008\pm0.004$ \\
 & $\chi^{2}/d.o.f.$ & 15/19 \\
\hline
\end{tabular}
\end{table}

A quantitative comparison between the cosine model and a constant (null) model indicates that periodic modulation is statistically preferred. In the 2--3\,keV band, the joint cosine fit yields $\chi^2/d.o.f. = 14/19$, while the constant model gives $\chi^2/d.o.f. = 31/22$. An F-test between the two models gives $F = 7.53$ with $p = 0.0016$, indicating that the cosine model is significantly better. This combined analysis indicates that both Stokes parameters exhibit coherent periodic modulation at the known orbital period.
For the 3--8\,keV band, the same test yields a better $\chi^2$, but no highly significant preference, for the cosine model: the joint fit gives $\chi^2/d.o.f. = 16/19$ versus $\chi^2/d.o.f. = 21/22$ for constant model, resulting in $F = 1.89$ with $p = 0.17 {>} 0.05$. Thus, the constant model cannot be fully rejected in this harder band.
In the full 2--8\,keV band, the cosine model again strongly outperforms the constant model, with $\chi^2/d.o.f. = 15/19$ against $\chi^2/d.o.f. = 39/22$. The F-test gives $F = 10.0$ with $p = 0.0004$. These results suggest that the overall 2--8\,keV modulation is primarily driven by the lower-energy photons. Similar results are obtained for the folded data reported in Appendix~\ref{app:Folded_QU}.

The fit yields a maximum PD at a phase $\phi = 0.22$ in 2--3\,keV, which coincides with the inferior conjunction phase. This phase alignment supports the interpretation that the observed polarization variability arises from orbital-phase-dependent scattering geometry.

Considering that the cosine fit provides enhanced polarization during the inferior conjunction, we performed a detailed polarimetric analysis at this orbital phase and out of it (non-inferior conjunction). The polarization degree (PD) and polarization angle (PA) are measured to be $(3.4\pm0.8)\%$ and $(50\pm7)^\circ$ in the 2--8\,keV band $(4.2\pm1.1)\%$ and $(46\pm8)^\circ$ in the 2--3\,keV band at inferior conjunction, while all PDs are below MDP$_{99}$ at non-inferior conjunction (see Table~\ref{tab:Polarization_results}).

\begin{table*}[!htb]
\centering
\caption{Polarization properties of inferior conjunction and non-inferior conjunction. Errors are reported at 68\%\,CL. Upper limits are reported at 90\%\,CL.}
\label{tab:Polarization_results}
\begin{tabular}{l ccc ccc}
\hline \\[-1.7ex]
 & \multicolumn{3}{c}{Inferior Conjunction} & \multicolumn{3}{c}{Non-Inferior Conjunction} \\ 
\cline{2-4} \cline{5-7} \\[-1.7ex]
 & 2--3\,keV & 3--8\,keV & 2--8\,keV & 2--3\,keV & 3--8\,keV & 2--8\,keV \\
\hline \\[-1.7ex]
$U/I$ (\%) & $4.2\pm1.1$ & $1.2\pm1.1$ & $3.4\pm0.8$ & $-0.1\pm0.6$ & $0.4\pm0.6$ & $0.1\pm0.4$ \\
$Q/I$ (\%) & $-0.1\pm1.1$ & $-1.6\pm1.1$ & $-0.5\pm0.8$ & $0.1\pm0.6$ & $-0.2\pm0.6$ & $0.0\pm0.4$ \\
PA ($^\circ$) & $46\pm8$ & $71\pm16$ & $50\pm7$ & -- & -- & -- \\
PD (\%) & $4.2\pm1.1$ & $2.0\pm1.1$ & $3.4\pm0.8$ & $<0.9$ & $<1.2$ & $<0.6$ \\
MDP$_{99}$ (\%) & 3.4 & 3.3 & 2.4 & 1.8 & 1.8 & 1.3 \\
\hline
\end{tabular}
\end{table*}
Interestingly, the PA in the 2--3\,keV band shows consistency with the orientation of the ionization cone (${\sim}45^\circ$) reported by \citet{Cooke2008}, which is interpreted as perpendicular to the accretion disk plane and aligned with the inferred jet direction. During the non-inferior conjunction phases, the polarization in both bands remains below the MDP$_{99}$.

\subsection{Spectral analysis} \label{subsec:Spectral analysis}

\begin{figure}[!ht]
\centering
\includegraphics[width=\linewidth]{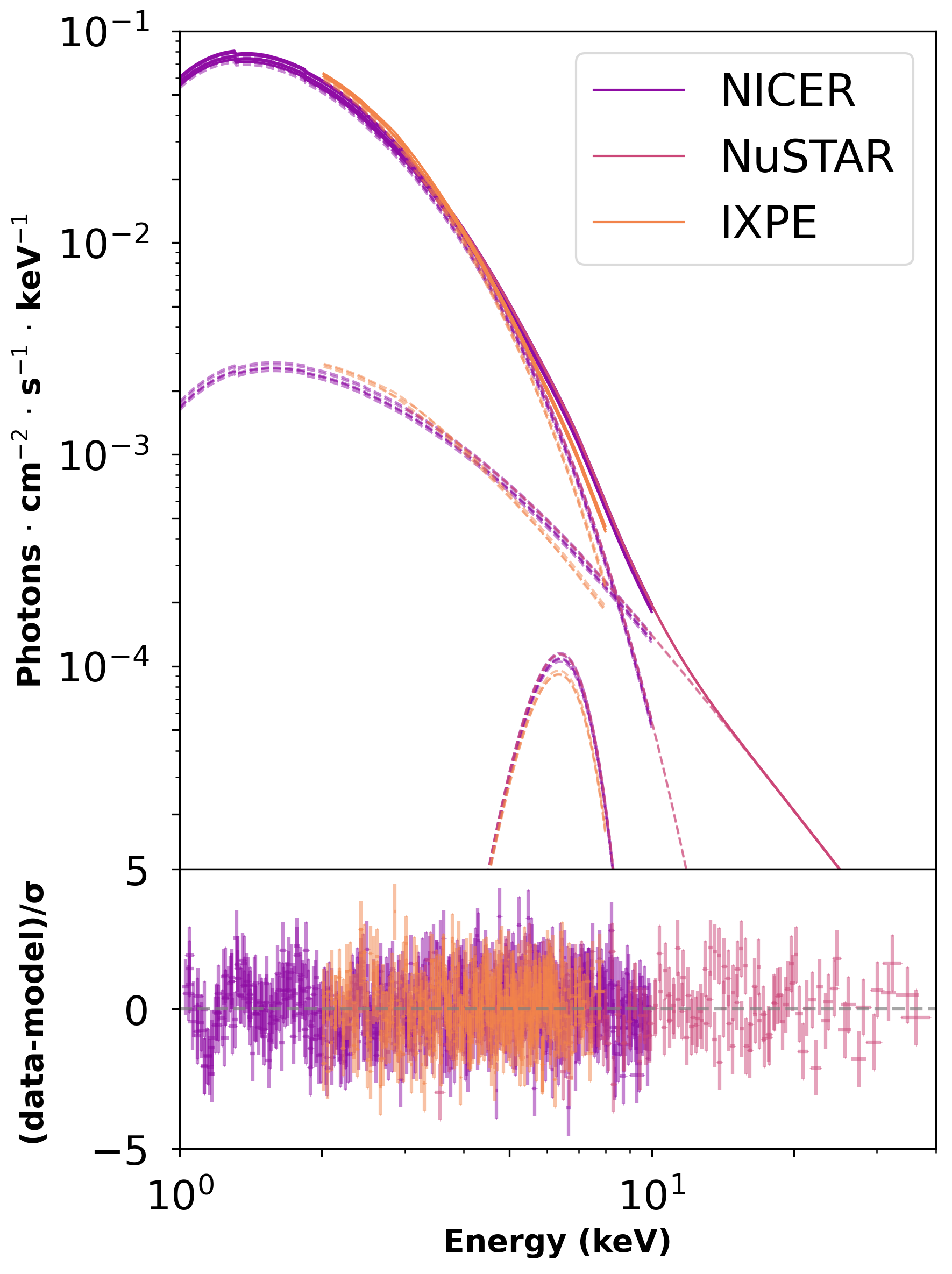}
\caption{X-ray spectra of LMC~X-1 in the inferior conjunction periods. Top: decomposition of the total model into individual components for each detector. Bottom: residuals in units of $\sigma$.}
\label{fig:spectra}
\end{figure}

We performed a joint spectral fit of the simultaneous \nicer, \nustar, and \ixpe data around the inferior conjunction phase.
Specifically, for NICER, only observations 5100070101, 5100070102, 5100070105, 5100070106, 5100070109, and 5100070110 overlap with the inferior conjunction phase.
Following the approach reported in \citet{Podgorny2023}, we used the model \texttt{MBPO * TBABS * (KERRBB + NTHCOMP + GAUSS)} in \textsc{XSPEC}. The \texttt{MBPO} is a multiplicative cross-calibration model used to reconcile discrepancies among the different instruments; its best-fit parameters are detailed in Appendix~\ref{app:mbpo}. The \texttt{KERRBB}\footnote{Following the approach of \citealt{Podgorny2023}, we fixed the normalization of the \texttt{KERRBB} component to unity, as this parameter is physically determined by the accretion rate and distance, and is not a free parameter.} component describes the general relativistic, multi-temperature blackbody radiation from a thin accretion disk around a Kerr-spinning BH, while \texttt{NTHCOMP} models the Comptonized emission from a hot corona; we also include a broad Gaussian line at 6.4\,keV due to a reflection component. The spectral parameters, as obtained from the best-fit, are shown in Table~\ref{tab:Best-fit_Parameters}, and the spectra and residuals are presented in Figure~\ref{fig:spectra}. In the 2--8\,keV band, the disk component contributes ${\sim}90\%$ of the total flux, confirming that LMC~X‑1 was firmly in the high/soft state during observation. 

\begin{table}[!htb]
\centering
\caption{Best-fit parameters for the spectral model as obtained at the inferior conjunction. Parameters in squared brackets are fixed. Errors are reported at 68\%\,CL. Fluxes are calculated in the 2--8\,keV energy band; in 2--3\,keV we have $1.6\times10^{-10}\,\mathrm{erg\,cm^{-2}\,s^{-1}}$ with 95\% of flux from \texttt{KERRBB} and 5\% from \texttt{NTHCOMP}; in 3--8\,keV we observe a flux of $2.1\times10^{-10}\,\mathrm{erg\,cm^{-2}\,s^{-1}}$ with 86\% of contribution from \texttt{KERRBB}, 13\% from \texttt{NTHCOMP} and the remaining 1\% from the \texttt{GAUSS}. The inter-instrument calibration is instead handled by the \texttt{MBPO} model (see Appendix~\ref{app:mbpo}).}
\label{tab:Best-fit_Parameters}
\begin{tabular}{l c c}
\hline
Component & Parameter (unit) & Inferior Conjunction \\
\hline
TBABS & $N_{\text{H}}$ ($10^{22}$ cm$^{-2}$) & $0.714\pm0.003$ \\[0.5em]
KERRBB & $\eta$ & [0] \\
 & $a$ & [0.92] \\
 & $i$ (deg) & [36.4] \\
 & $M_{\rm BH}$ ($M_{\odot}$) & [10.9] \\
 & $M_{\rm dd}$ ($10^{18}$\,g\,s$^{-1}$) & $1.57\pm0.02$ \\
 & $D_{\rm BH}$ (kpc) & [50] \\
 & $hd$ & [1.7] \\
 & $r_{\rm flag}$ & [1] \\
 & $l_{\rm flag}$ & [0] \\
 & norm & [1] \\[0.5em]
NTHCOMP & $\Gamma$ & $2.67\pm0.05$ \\
 & $kT_{\rm e}$ (keV) & [100] \\
 & $kT_{\rm bb}$ (keV) & $1.3^{+0.3}_{-0.2}$ \\
 & norm ($10^{-3}$) & $6\pm2$ \\[0.5em]
GAUSS & $E$ (keV) & [6.4] \\
 & $\sigma$ (keV) & $0.87\pm0.09$ \\
 & norm ($\times10^{-4}$) & $2.5\pm0.4$ \\[0.5em]
\hline
$\chi^{2}$/d.o.f. & & 1239/1096 \\
\hline
Flux$_{\mathrm{total}}$ & ($10^{-10}\,\mathrm{erg\,cm^{-2}\,s^{-1}}$) & 3.739 \\
Flux$_{\mathrm{kerrbb}}$ & ($10^{-10}\,\mathrm{erg\,cm^{-2}\,s^{-1}}$) & 3.364 \\
Flux$_{\mathrm{nthcomp}}$ & ($10^{-10}\,\mathrm{erg\,cm^{-2}\,s^{-1}}$) & 0.350 \\
Flux$_{\mathrm{gauss}}$ & ($10^{-10}\,\mathrm{erg\,cm^{-2}\,s^{-1}}$) & 0.025 \\
\hline
\end{tabular}
\end{table}

\subsection{Spectro-polarimetric analysis} \label{subsec:Spectro-polarimetric analysis}

Based on the spectral model reported in Table~\ref{tab:Best-fit_Parameters}, we carried out a spectro-polarimetric analysis by jointly fitting the Stokes $I$, $Q$ and $U$ spectra from \ixpe in the 2--8\,keV energy band. The spectral parameters of the model \texttt{MBPO * TBABS * (KERRBB + NTHCOMP + GAUSS)} were fixed at their best-fit values, allowing only the polarimetric parameters to vary. 

Considering the higher PD in the low 2-3\,keV energy band with respect to 3--8\,keV, we attempted a spectro-polarimetric fit by using both \texttt{POLLIN} and \texttt{POLCONST} to the whole spectrum after assuming a non-polarized Gaussian. The latter assumption is due to the fact that we expect fluorescence lines to be almost unpolarized \citep{Churazov02, Veledina24, LaMonaca25}. We first applied the \texttt{POLLIN} model, which assumes a linear energy dependence of both the PD and PA \citep{DiMarco26}. The fit yielded $PD_\textrm{1\,keV} = (6\pm2)\%$ and $PA_\textrm{1\,keV} = (-2\pm15)^\circ$, with slopes of $(-2.2\pm0.9)\%\,\mathrm{keV}^{-1}$ and $(34\pm7)^\circ\,\mathrm{keV}^{-1}$ for the PD and PA, respectively ($\chi^2/\mathrm{d.o.f.}=584/523$). 
We also tried the \texttt{POLPOW} model, which assumes a power‐law energy dependence, but the PD at 1\,keV could not be constrained.
When replacing \texttt{POLLIN} with a simple constant polarization model, \texttt{POLCONST}, we obtained a PD of $(2.0\pm0.6)\%$ and a PA of $(55\pm9)^{\circ}$ ($\chi^2/\text{d.o.f.}$=587/525). The F-test reports a value of 1.34 (P$=26$\%) indicating no significant improvement with \texttt{POLLIN} over \texttt{POLCONST}. The \texttt{POLCONST} values are consistent within uncertainties with the model-independent results; furthermore, the analysis of the 2--3\,keV band yields a higher PD of $(3.0\pm1.1)\%$ with $\mathrm{PA} = (47\pm11)^{\circ}$ ($\chi^2/\mathrm{d.o.f.} = 129/94$), suggesting that the polarization, if present, may be more prominent at lower energies, as hinted by the negative slope in the \texttt{POLLIN} best-fit result. As a result, we find that the larger PD at lower energy is not a simple linear variation but rather a more complex trend.

To investigate the origin of polarization across different spectral components, we performed a polarimetric analysis by convolving each spectral component with \texttt{POLCONST}. The results are summarized in Table~\ref{tab:Best-fit_polarization}. When polarization is free to vary for both the disk and corona components, we obtained an unconstrained polarization for the corona and polarization for the disk component, which is dominant. We therefore explored two scenarios: firstly, assuming that all the polarization arises from the \texttt{KERRBB} component, which dominates the whole spectrum but is diluted by Comptonization at higher energies; then, assuming that it arises from the \texttt{NTHCOMP} component. In the scenario where the polarization is attributed exclusively to the Comptonization in the corona, the resulting polarization degree is poorly determined and unphysical, with values ${\geq 20}\%$; this is due to the low relative flux of this spectral component, which produces in \textsc{XSPEC} these large values of polarization. By contrast, when the polarization is assumed to originate from the disk component, the PD is consistent with the model-independent measurements, where a hint of polarization is observed preferentially in the lower energy band. From a statistical point of view, the $\chi^2$ provides no preference between the different scenarios. These results, although not conclusive, suggest that the observed polarization in LMC~X-1 is more likely associated with the disk emission than the corona, and the lower polarization at higher energies could be due to depolarization and dilution. To further investigate this dilution effect, we adopted the dilution formula from \citet{Di_Marco2023} and applied it to our two‑component model (a polarized disk plus an unpolarized corona and a Gaussian component), using the flux fractions of the disk in each energy band derived from our spectral fitting. 
Specifically, we computed the expected net polarization as $P = P_{disk} \times F_{diskb} / F_{total}$. For the \texttt{PCUBE} analysis, we can determine from the relative fluxes the disk polarization in different energy bins: $P_{disk} (\textrm{2--3\,keV}) = P \times F_{total}/F_{diskb} = (4.4\pm1.2)\%/0.95 =(4.6\pm1.3)\%$; $P_{disk} (\textrm{3--8\,keV}) =
(2.3\pm1.3)\%$; $P_{disk} (\textrm{2--8\,keV}) = (3.8\pm0.9)\%$. Despite the large uncertainties, the results point to a dilution effect, with intrinsic $P_{disk}$ values determined in the different energy bands compatible within ${\leq}1.4\sigma$.

\begin{table}[!htb]
\centering
\footnotesize
\caption{Spectro-polarimetric results for the inferior conjunction periods using the spectral model in Table~\ref{tab:Best-fit_Parameters}. The gaussian component is assumed to be unpolarized. All uncertainties are reported at the 68\%\,CL, upper limits are at 90\%\,CL.}
\label{tab:Best-fit_polarization}
\begin{tabular}{l c c c}
\hline
 & Parameter (unit) & 2--3\,keV & 2--8\,keV \\
\hline \\[-1.7ex]
POLCONST & $PD_1$ (\%) & $8.3^{+1.1}_{-4.0}$ & $3.4\pm1.7$ \\
(KERRBB) & $PA_1$ ($^{\circ}$) & $40\pm20$ & $30\pm15$ \\[0.5em]
POLCONST & $PD_2$ (\%) & Unconstrained & $<56\%$ \\
(NTHCOMP) & $PA_2$ ($^{\circ}$) & -- & $-81\pm18$ \\[0.5em]
 & $\chi^{2}$/d.o.f. & 128/92 & 584/523 \\[0.5em]
\hline \\[-1.7ex]
POLCONST & $PD_1$ (\%) & $3.0\pm1.1$ & $2.2\pm0.7$ \\
(KERRBB) & $PA_1$ ($^{\circ}$) & $47\pm11$ & $55\pm9$ \\[0.5em]
POLCONST & $PD_2$ (\%) & -- & -- \\
(NTHCOMP) & $PA_2$ ($^{\circ}$) & -- & -- \\[0.5em]
 & $\chi^{2}$/d.o.f. & 129/94 & 587/525 \\[0.5em]
\hline \\[-1.7ex]
POLCONST & $PD_1$ (\%) & -- & -- \\
(KERRBB) & $PA_1$ ($^{\circ}$) & -- & -- \\[0.5em]
POLCONST & $PD_2$ (\%) & $51\pm20$ & $25\pm8$ \\
(NTHCOMP) & $PA_2$ ($^{\circ}$) & $47\pm11$ & $62\pm10$ \\[0.5em]
 & $\chi^{2}$/d.o.f. & 130/94 & 588/525 \\[0.5em]
\hline
\end{tabular}
\end{table}

\section{Discussion} \label{sec:Discussion}

The October 2022 data show evidence for an orbital-phase dependence of the X‑ray polarization from LMC~X‑1. The PD peaks around inferior conjunction, while it remains below the MDP$_{99}$ during non‑inferior conjunction phases.
This is not inconsistent with the low, orbit‑averaged polarization reported by \citet{Podgorny2023}; rather, it uncovers the phase‑dependent variability that had been hidden by the averaged measurement. 
The 2--3\,keV Stokes parameters of LMC~X-1 are modulated at the orbital period ($\chi^2/d.o.f. = 14/19$ against $31/22$ for a constant; $F = 7.53$, $p = 0.0016$), while the 3--8\,keV band shows no significant modulation ($F = 1.89$, $p = 0.166$). The 2--8\,keV band, which is dominated by soft photons, follows the 2--3\,keV behavior ($F = 10.03$, $p = 0.0004$). We therefore treat the soft excess of the modulation as the primary diagnostic throughout this section.

Two features of this measurement challenge the standard interpretation of the high/soft state, in which the polarization is produced by electron scattering in the optically thick disk atmosphere. For $i = 36.4^{\circ}$, such an atmosphere yields PD~$\sim1.3\%$ \citep{Chandrasekhar60}, the value invoked by \citet{Podgorny2023} to account for their orbit-averaged upper limit, whereas PD${\sim}4\%$ is measured at inferior conjunction. Second, the polarization angle ($46^{\circ} \pm 8^{\circ}$) is consistent with the position angle of the ionization cone (${\sim}45^{\circ}$), which is presumed to trace the disk normal and the inferred jet axis. In contrast, an optically thick disk scattering atmosphere produces polarization perpendicular to the disk normal (i.e., ${\sim}135^{\circ}$). Thus, any explanation of the measured polarization must provide both more polarized emission and a polarization angle along the inferred jet axis. Accordingly, we propose two models below. In the first, the polarization is the sum of two components with fixed position angles, whose relative contribution changes along the orbit. In the second model, the orbital variation is produced by the Faraday rotation broken symmetry along the line of sight.

\subsection{Geometry of the orbital modulation} \label{Geometry}

We model the phase dependence with the cosine formulation of \citet{Kravtsov2025}, generalized to allow unequal amplitudes and a $\pi$ phase difference between $Q/I$ and $U/I$, i.e., Eq.~\ref{eq:QUcos}. The first result is that the polarization degree modulates with amplitude $A_{2-3{\rm \,keV}} = (3.2\pm0.8)\%$ in 2-3\,keV, with a constant polarization component that is stable along the orbit of $|C|_{2-3{\rm \,keV}} = (0.5\pm0.5)\%$. The modulated component shows a PD amplitude that depends on energy: $A_{3-8{\rm \,keV}} = (1.6\pm0.7)\%$ and $A_{2-8{\rm \,keV}} = (2.7\pm0.6)\%$. Whereas, the unmodulated PD component is energy independent: $|C|_{3-8{\rm \,keV}}=(0.7\pm0.5)\%$ and $|C|_{2-8{\rm \,keV}}=(0.7\pm0.4)\%$. These energy dependencies point to two nearly orthogonal polarized components that cancel when their contributions are equal, with their mixing varying along the orbit. Furthermore, the amplitude, and consequently the component that is enhancing the polarization at low energy, is strongly dependent on the orbital phase.

Thus, assuming the two components with nearly orthogonal angles are (i) an optically thin scattering medium, whose polarization is parallel to the disk normal, such as the ionization cone, and (ii) an optically thick disk, whose polarization is perpendicular to the disk normal. The two components arise in spatially distinct regions. The optically thin component is a scattering region above the orbital plane, possibly the polar outflow traced by the ionization cone \citep{Cooke2008}, which extends beyond the densest part of the companion's occulting medium and remains largely unaffected by absorption in inferior conjunction. On the other hand, the optically thick disk component is direct disk emission that emerges along the orbital plane and must pass through the full column of the companion's wind and the bulge, so it is partially absorbed near inferior conjunction. In conclusion, this scenario has absorption along the line of sight from the companion's wind, which partially absorbs emission from the accretion disk, while scattering in the ionization cone, unaffected by absorption, produces the raised PD dominated by PA along the disk normal during inferior conjunction. In this scenario, the direct disk emission provides a dilution of the polarization due to absorption, and when the companion is along the line of sight, the direct component is more absorbed at low energy, which reduces the dilution effect below 3\,keV, leaving it almost unchanged above 3\,keV.

\subsection{Faraday rotation in a magnetized outflow} \label{faraday}

In the second model, the polarized photons are produced by the accretion disk magnetic field due to Faraday rotation. A linearly polarized wave can be split into two equal circular components (left- and right-handed), which interact differently with the magnetic field in the presence of electrons. As a result, they travel at different speeds in the magnetized plasma, causing the polarization plane to rotate as the X-rays emerge from the plasma \citep{Barnier2024}. The polarization is therefore produced by accretion disk photons, and Faraday rotation along the disk plane is the source of the orbital modulation. Following \citet{Barnier2024}, Faraday rotation can produce nonuniform polarization along the disk. For example, a toroidal magnetic field causes sign changes between the left and right halves of the disk along the line of sight, while the front and back sides show no change. On large scales, this effect produces polarization that depends on the observer's inclination. Assuming an ordered magnetic field in LMC~X-1, the companion star's position along the orbit changes the occultation of the accretion disk along the line of sight, and we observe changing polarization due to the asymmetry induced by Faraday rotation, which also becomes visible on large scales. The phenomenological fit obtained by Eq.~\ref{eq:QUcos} provides the amplitude of the modulated component, which we attribute to the maximum polarization due to Faraday rotation, and hence constrains the ordered magnetic field along the line of sight, $B_{\rm LoS}$. Following \citet{Barnier2024}:
\begin{equation}\label{eq:B}
\Delta\varphi
= 5^{\circ}\!.5\left(\frac{B_{\rm LoS}}{10^{6}\,{\rm G}}\right)
\left(\frac{2.5\,{\rm keV}}{E}\right)^{2}\tau_{\rm LoS},
\end{equation}
where $\Delta\varphi$ is the polarization shift induced by Faraday rotation, $\tau_{\rm LoS}$ is the optical depth along the line of sight (hereafter assumed equal to 1), and $E$ is the mean energy (i.e., the average $E$ accounting for the number of photons as a function of energy detected by \ixpe). In Eq.~\ref{eq:B}, we include: $E= 2.38$\,keV for 2--3\,keV and 3.55\,keV for 3--8\,keV, and a maximum of PA variations between the modulated and the unmodulated components of $\Delta\varphi=19^{\circ}\pm25^{\circ}$ in 2--3\,keV and $0^{\circ}\pm15^{\circ}$ in 3--8\,keV. The result provides a $90\%$ CL upper limit: $B_{\rm LoS}<9.7\times10^{6}$\,G.

Besides the position angle, the magnetic field intensity can also be constrained from the amplitude of the modulated component. The largest PD, $P_{\max}$, is the intrinsic polarization $p_0$ before the Faraday rotation; the amplitude of the modulation then gives:
\begin{equation}\label{eq:Amod}
A=\sqrt{A_Q^2+A_U^2},\qquad \sin\Delta\varphi=\frac{A}{2P_{\max}}.
\end{equation}
In the 2--3\,keV band $A=(3.2\pm0.8)\%$ and $P_{\max}=(3.6\pm0.9)\%$; in 3--8\,keV, $A=(1.6\pm0.7)\%$ and $P_{\max}=(2.3\pm0.8)\%$. These numbers provide a $90\%$ CL upper limit: $|B_{\rm LoS}|<9.7\times10^{6}$\,G, consistent with those obtained from the previous approach. This estimate does not depend on the orientation of the constant term, which is poorly constrained, and therefore confirms independently the upper limit we obtain.

Faraday rotation can explain the modulation of polarization along the orbit, but not the large PD we measured; for the latter, we need a second polarized component that sums or cancels with the effects of Faraday rotation. The \ixpe sample of high/soft sources is still too small to draw any conclusions about their polarization in terms of Faraday rotation. Among those with known radio jet directions, only Cyg~X-1 shows a high/soft-state PA parallel to the jet, and its polarization is likewise attributed to the outflow above the disk \citep{Steiner2024, Krawczynski2025}. For the remaining sources, the jet direction is unknown, or no significant polarization has been detected in the high/soft state, so the prediction remains untested.

\section{Conclusions} \label{sec:conclusions}

We have performed a comprehensive phase-resolved analysis of the black hole X-ray binary LMC~X-1 using simultaneous \ixpe, \nicer, and \nustar observations from the 2022 October campaign. Our analysis shows evidence for an orbital-phase dependence of the X-ray polarization, with a PD of $(4.2\pm1.1)\%$ detected in the 2--3\,keV band during the inferior conjunction. The PA in this orbital phase of $(46\pm8)^{\circ}$ aligns with the ionization cone and the inferred jet direction, and our spectro-polarimetric fits further identify the disk component as the origin of the observed polarization, disfavoring the corona as the dominant polarization source. The polarization remains below the MDP$_{99}$ during non-inferior conjunction phases, revealing a phase-dependent variability that was previously hidden in the orbit-averaged measurement.

Modeling the Stokes parameters with a cosine modulation, an amplitude of $(3.15\pm0.76)\%$ is measured, hinting at the presence of two polarized components with nearly orthogonal PAs. We propose two frameworks for explaining the results. In the first, an optically thin scattering above the orbital plane, polarized parallel to the disk normal and the inferred jet, enhances the polarization at inferior conjunction when the accretion disk emission, polarized perpendicular to it, is absorbed by the companion's wind. In the second, the disk photons show a modulated polarization due to Faraday rotation induced by an ordered magnetic field; this effect provides a $90\%$ CL constraint on the magnetic field of $<9.7\times10^{6}$\,G. This model requires an additional polarized component to explain the high PD we measured.

Our results suggest that the soft-band polarization detected in LMC~X-1 arises from two components. Future X-ray polarimetry missions could improve our understanding of this system.

\begin{acknowledgements}
This work is supported by National Natural Science Foundation of China (grant No. 12422306), and Natural Science Foundation of Guangxi (grant No. 2025GXNSFDA02850001), and Bagui Scholars Program (XF). 
This work is also supported by the Guangxi Science and Technology Innovation Platform Program (Leitai Action Plan, Grant No. Guike LT2600640026), Guangxi Key R$\&$D Program (Guangxi Funeng Action Plan, Grant No. Guike FN2504240040), and the ``Guangxi Highland of Innovation Talents'' Program. ADM and FLM are supported by INAF AF2025 Research Grant ``SAGE: Studying the Accretion-flow GEometry in Weakly Magnetized Neutron Stars" (PI A. Di Marco).
\end{acknowledgements}

\bibliographystyle{aa}
\bibliography{refs}

@INPROCEEDINGS{Arnaud1996,
       author = {{Arnaud}, K.~A.},
        title = "{XSPEC: The First Ten Years}",
    booktitle = {Astronomical Data Analysis Software and Systems V},
         year = 1996,
       editor = {{Jacoby}, George H. and {Barnes}, Jeannette},
       series = {Astronomical Society of the Pacific Conference Series},
       volume = {101},
        month = jan,
        pages = {17},
       adsurl = {https://ui.adsabs.harvard.edu/abs/1996ASPC..101...17A}
}

@INPROCEEDINGS{Arzoumanian2014,
       author = {{Arzoumanian}, Z. and {Gendreau}, K.~C. and {Baker}, C.~L. and {Cazeau}, T. and {Hestnes}, P. and {Kellogg}, J.~W. and {Kenyon}, S.~J. and {Kozon}, R.~P. and {Liu}, K.-C. and {Manthripragada}, S.~S. and {Markwardt}, C.~B. and {Mitchell}, A.~L. and {Mitchell}, J.~W. and {Monroe}, C.~A. and {Okajima}, T. and {Pollard}, S.~E. and {Powers}, D.~F. and {Savadkin}, B.~J. and {Winternitz}, L.~B. and {Chen}, P.~T. and {Wright}, M.~R. and {Foster}, R. and {Prigozhin}, G. and {Remillard}, R. and {Doty}, J.},
        title = "{The neutron star interior composition explorer (NICER): mission definition}",
    booktitle = {Space Telescopes and Instrumentation 2014: Ultraviolet to Gamma Ray},
         year = 2014,
       editor = {{Takahashi}, Tadayuki and {den Herder}, Jan-Willem A. and {Bautz}, Mark},
       series = {Society of Photo-Optical Instrumentation Engineers (SPIE) Conference Series},
       volume = {9144},
        month = jul,
          eid = {914420},
        pages = {914420},
          doi = {10.1117/12.2056811},
       adsurl = {https://ui.adsabs.harvard.edu/abs/2014SPIE.9144E..20A}
}

@ARTICLE{Baldini2022,
       author = {{Baldini}, Luca and {Bucciantini}, Niccol{\`o} and {Lalla}, Niccol{\`o} Di and {Ehlert}, Steven and {Manfreda}, Alberto and {Negro}, Michela and {Omodei}, Nicola and {Pesce-Rollins}, Melissa and {Sgr{\`o}}, Carmelo and {Silvestri}, Stefano},
        title = "{ixpeobssim: A simulation and analysis framework for the imaging X-ray polarimetry explorer}",
      journal = {SoftwareX},
         year = 2022,
        month = jul,
       volume = {19},
          eid = {101194},
        pages = {101194},
          doi = {10.1016/j.softx.2022.101194},
archivePrefix = {arXiv},
       eprint = {2203.06384},
 primaryClass = {astro-ph.IM},
       adsurl = {https://ui.adsabs.harvard.edu/abs/2022SoftX..1901194B}
}

@ARTICLE{Barnier2024,
       author = {{Barnier}, Samuel and {Done}, Chris},
        title = "{Making the Invisible Visible: Magnetic Fields in Accretion Flows Revealed by X-Ray Polarization}",
      journal = {\apj},
         year = 2024,
        month = dec,
       volume = {977},
       number = {2},
          eid = {201},
        pages = {201},
          doi = {10.3847/1538-4357/ad9277},
archivePrefix = {arXiv},
       eprint = {2404.12815},
 primaryClass = {astro-ph.HE},
       adsurl = {https://ui.adsabs.harvard.edu/abs/2024ApJ...977..201B}
}

@ARTICLE{Bhuvana2021,
       author = {{Bhuvana}, G.~R. and {Radhika}, D. and {Agrawal}, V.~K. and {Mandal}, S. and {Nandi}, A.},
        title = "{Broad-band 'spectro-temporal' features of extragalactic black hole binaries LMC X-1 and LMC X-3: an AstroSat perspective}",
      journal = {\mnras},
         year = 2021,
        month = mar,
       volume = {501},
       number = {4},
        pages = {5457-5467},
          doi = {10.1093/mnras/staa4012},
archivePrefix = {arXiv},
       eprint = {2012.11918},
 primaryClass = {astro-ph.HE},
       adsurl = {https://ui.adsabs.harvard.edu/abs/2021MNRAS.501.5457B}
}

@BOOK{Chandrasekhar60,
       author = {{Chandrasekhar}, Subrahmanyan},
        title = "{Radiative transfer}",
         year = 1960,
       adsurl = {https://ui.adsabs.harvard.edu/abs/1960ratr.book.....C}
}

@ARTICLE{Churazov02,
       author = {{Churazov}, E. and {Sunyaev}, R. and {Sazonov}, S.},
        title = "{Polarization of X-ray emission from the Sgr B2 cloud}",
      journal = {\mnras},
         year = 2002,
        month = mar,
       volume = {330},
       number = {4},
        pages = {817-820},
          doi = {10.1046/j.1365-8711.2002.05113.x},
archivePrefix = {arXiv},
       eprint = {astro-ph/0111065},
 primaryClass = {astro-ph},
       adsurl = {https://ui.adsabs.harvard.edu/abs/2002MNRAS.330..817C}
}

@ARTICLE{Cooke2007,
       author = {{Cooke}, Ryan and {Kuncic}, Zdenka and {Sharp}, Rob and {Bland-Hawthorn}, Joss},
        title = "{Spectacular Trailing Streamers near LMC X-1: The First Evidence of a Jet?}",
      journal = {\apjl},
         year = 2007,
        month = oct,
       volume = {667},
       number = {2},
        pages = {L163-L166},
          doi = {10.1086/522517},
archivePrefix = {arXiv},
       eprint = {0708.2314},
 primaryClass = {astro-ph},
       adsurl = {https://ui.adsabs.harvard.edu/abs/2007ApJ...667L.163C}
}

@ARTICLE{Cooke2008,
       author = {{Cooke}, R. and {Bland-Hawthorn}, J. and {Sharp}, R. and {Kuncic}, Z.},
        title = "{Ionization Cone in the X-Ray Binary LMC X-1}",
      journal = {\apjl},
         year = 2008,
        month = nov,
       volume = {687},
       number = {1},
        pages = {L29},
          doi = {10.1086/593169},
archivePrefix = {arXiv},
       eprint = {0809.2140},
 primaryClass = {astro-ph},
       adsurl = {https://ui.adsabs.harvard.edu/abs/2008ApJ...687L..29C}
}

@ARTICLE{DiMarco22a,
       author = {{Di Marco}, Alessandro and {Costa}, Enrico and {Muleri}, Fabio and {Soffitta}, Paolo and {Fabiani}, Sergio and {La Monaca}, Fabio and {Rankin}, John and {Xie}, Fei and {Bachetti}, Matteo and {Baldini}, Luca and {Baumgartner}, Wayne and {Bellazzini}, Ronaldo and {Brez}, Alessandro and {Castellano}, Simone and {Del Monte}, Ettore and {Di Lalla}, Niccol{\`o} and {Ferrazzoli}, Riccardo and {Latronico}, Luca and {Maldera}, Simone and {Manfreda}, Alberto and {O'Dell}, Stephen L. and {Perri}, Matteo and {Pesce-Rollins}, Melissa and {Puccetti}, Simonetta and {Ramsey}, Brian D. and {Ratheesh}, Ajay and {Sgr{\`o}}, Carmelo and {Spandre}, Gloria and {Tennant}, Allyn F. and {Tobia}, Antonino and {Trois}, Alessio and {Weisskopf}, Martin C.},
        title = "{A Weighted Analysis to Improve the X-Ray Polarization Sensitivity of the Imaging X-ray Polarimetry Explorer}",
      journal = {\aj},
         year = 2022,
        month = apr,
       volume = {163},
       number = {4},
          eid = {170},
        pages = {170},
          doi = {10.3847/1538-3881/ac51c9},
archivePrefix = {arXiv},
       eprint = {2202.01093},
 primaryClass = {astro-ph.IM},
       adsurl = {https://ui.adsabs.harvard.edu/abs/2022AJ....163..170D}
}

@ARTICLE{Di_Marco2023,
       author = {{Di Marco}, Alessandro and {Soffitta}, Paolo and {Costa}, Enrico and {Ferrazzoli}, Riccardo and {La Monaca}, Fabio and {Rankin}, John and {Ratheesh}, Ajay and {Xie}, Fei and {Baldini}, Luca and {Del Monte}, Ettore and {Ehlert}, Steven R. and {Fabiani}, Sergio and {Kim}, Dawoon E. and {Muleri}, Fabio and {O'Dell}, Stephen L. and {Ramsey}, Brian D. and {Rubini}, Alda and {Sgr{\`o}}, Carmelo and {Silvestri}, Stefano and {Tennant}, Allyn F. and {Weisskopf}, Martin C.},
        title = "{Handling the Background in IXPE Polarimetric Data}",
      journal = {\aj},
         year = 2023,
        month = apr,
       volume = {165},
       number = {4},
          eid = {143},
        pages = {143},
          doi = {10.3847/1538-3881/acba0f},
archivePrefix = {arXiv},
       eprint = {2302.02927},
 primaryClass = {astro-ph.IM},
       adsurl = {https://ui.adsabs.harvard.edu/abs/2023AJ....165..143D}
}

@ARTICLE{DiMarco25,
       author = {{Di Marco}, Alessandro and {La Monaca}, Fabio and {Bobrikova}, Anna and {Stella}, Luigi and {Papitto}, Alessandro and {Poutanen}, Juri and {Baglio}, Maria Cristina and {Bachetti}, Matteo and {Loktev}, Vladislav and {Pilia}, Maura and {Rogantini}, Daniele},
        title = "{X-Ray Dips and Polarization Angle Swings in GX 13+1}",
      journal = {\apjl},
         year = 2025,
        month = feb,
       volume = {979},
       number = {2},
          eid = {L47},
        pages = {L47},
          doi = {10.3847/2041-8213/ada7f8},
archivePrefix = {arXiv},
       eprint = {2501.05511},
 primaryClass = {astro-ph.HE},
       adsurl = {https://ui.adsabs.harvard.edu/abs/2025ApJ...979L..47D}
}

@ARTICLE{DiMarco26,
       author = {{Di Marco}, Alessandro},
        title = "{The hitchhiker's guide to the IXPE data analysis}",
      journal = {arXiv e-prints},
         year = 2026,
        month = apr,
          eid = {arXiv:2604.03366},
        pages = {arXiv:2604.03366},
          doi = {10.48550/arXiv.2604.03366},
archivePrefix = {arXiv},
       eprint = {2604.03366},
 primaryClass = {astro-ph.HE},
       adsurl = {https://ui.adsabs.harvard.edu/abs/2026arXiv260403366D}
}

@INPROCEEDINGS{Gendreau2016,
       author = {{Gendreau}, Keith C. and {Arzoumanian}, Zaven and {Adkins}, Phillip W. and {Albert}, Cheryl L. and {Anders}, John F. and {Aylward}, Andrew T. and {Baker}, Charles L. and {Balsamo}, Erin R. and {Bamford}, William A. and {Benegalrao}, Suyog S. and {Berry}, Daniel L. and {Bhalwani}, Shiraz and {Black}, J. Kevin and {Blaurock}, Carl and {Bronke}, Ginger M. and {Brown}, Gary L. and {Budinoff}, Jason G. and {Cantwell}, Jeffrey D. and {Cazeau}, Thoniel and {Chen}, Philip T. and {Clement}, Thomas G. and {Colangelo}, Andrew T. and {Coleman}, Jerry S. and {Coopersmith}, Jonathan D. and {Dehaven}, William E. and {Doty}, John P. and {Egan}, Mark D. and {Enoto}, Teruaki and {Fan}, Terry W. and {Ferro}, Deneen M. and {Foster}, Richard and {Galassi}, Nicholas M. and {Gallo}, Luis D. and {Green}, Chris M. and {Grosh}, Dave and {Ha}, Kong Q. and {Hasouneh}, Monther A. and {Heefner}, Kristofer B. and {Hestnes}, Phyllis and {Hoge}, Lisa J. and {Jacobs}, Tawanda M. and {J{\o}rgensen}, John L. and {Kaiser}, Michael A. and {Kellogg}, James W. and {Kenyon}, Steven J. and {Koenecke}, Richard G. and {Kozon}, Robert P. and {LaMarr}, Beverly and {Lambertson}, Mike D. and {Larson}, Anne M. and {Lentine}, Steven and {Lewis}, Jesse H. and {Lilly}, Michael G. and {Liu}, Kuochia Alice and {Malonis}, Andrew and {Manthripragada}, Sridhar S. and {Markwardt}, Craig B. and {Matonak}, Bryan D. and {Mcginnis}, Isaac E. and {Miller}, Roger L. and {Mitchell}, Alissa L. and {Mitchell}, Jason W. and {Mohammed}, Jelila S. and {Monroe}, Charles A. and {Montt de Garcia}, Kristina M. and {Mul{\'e}}, Peter D. and {Nagao}, Louis T. and {Ngo}, Son N. and {Norris}, Eric D. and {Norwood}, Dwight A. and {Novotka}, Joseph and {Okajima}, Takashi and {Olsen}, Lawrence G. and {Onyeachu}, Chimaobi O. and {Orosco}, Henry Y. and {Peterson}, Jacqualine R. and {Pevear}, Kristina N. and {Pham}, Karen K. and {Pollard}, Sue E. and {Pope}, John S. and {Powers}, Daniel F. and {Powers}, Charles E. and {Price}, Samuel R. and {Prigozhin}, Gregory Y. and {Ramirez}, Julian B. and {Reid}, Winston J. and {Remillard}, Ronald A. and {Rogstad}, Eric M. and {Rosecrans}, Glenn P. and {Rowe}, John N. and {Sager}, Jennifer A. and {Sanders}, Claude A. and {Savadkin}, Bruce and {Saylor}, Maxine R. and {Schaeffer}, Alexander F. and {Schweiss}, Nancy S. and {Semper}, Sean R. and {Serlemitsos}, Peter J. and {Shackelford}, Larry V. and {Soong}, Yang and {Struebel}, Jonathan and {Vezie}, Michael L. and {Villasenor}, Joel S. and {Winternitz}, Luke B. and {Wofford}, George I. and {Wright}, Michael R. and {Yang}, Mike Y. and {Yu}, Wayne H.},
        title = "{The Neutron star Interior Composition Explorer (NICER): design and development}",
    booktitle = {Space Telescopes and Instrumentation 2016: Ultraviolet to Gamma Ray},
         year = 2016,
       editor = {{den Herder}, Jan-Willem A. and {Takahashi}, Tadayuki and {Bautz}, Marshall},
       series = {Society of Photo-Optical Instrumentation Engineers (SPIE) Conference Series},
       volume = {9905},
        month = jul,
          eid = {99051H},
        pages = {99051H},
          doi = {10.1117/12.2231304},
       adsurl = {https://ui.adsabs.harvard.edu/abs/2016SPIE.9905E..1HG}
}

@ARTICLE{Gou2009,
       author = {{Gou}, Lijun and {McClintock}, Jeffrey E. and {Liu}, Jifeng and {Narayan}, Ramesh and {Steiner}, James F. and {Remillard}, Ronald A. and {Orosz}, Jerome A. and {Davis}, Shane W. and {Ebisawa}, Ken and {Schlegel}, Eric M.},
        title = "{A Determination of the Spin of the Black Hole Primary in LMC X-1}",
      journal = {\apj},
         year = 2009,
        month = aug,
       volume = {701},
       number = {2},
        pages = {1076-1090},
          doi = {10.1088/0004-637X/701/2/1076},
archivePrefix = {arXiv},
       eprint = {0901.0920},
 primaryClass = {astro-ph.HE},
       adsurl = {https://ui.adsabs.harvard.edu/abs/2009ApJ...701.1076G}
}

@ARTICLE{Hanke2010,
       author = {{Hanke}, M. and {Wilms}, J. and {Nowak}, M.~A. and {Barrag{\'a}n}, L. and {Schulz}, N.~S.},
        title = "{The column density towards LMC X-1}",
      journal = {\aap},
         year = 2010,
        month = jan,
       volume = {509},
          eid = {L8},
        pages = {L8},
          doi = {10.1051/0004-6361/200913583},
archivePrefix = {arXiv},
       eprint = {0912.3719},
 primaryClass = {astro-ph.HE},
       adsurl = {https://ui.adsabs.harvard.edu/abs/2010A&A...509L...8H}
}

@ARTICLE{Harrison2013,
       author = {{Harrison}, Fiona A. and {Craig}, William W. and {Christensen}, Finn E. and {Hailey}, Charles J. and {Zhang}, William W. and {Boggs}, Steven E. and {Stern}, Daniel and {Cook}, W. Rick and {Forster}, Karl and {Giommi}, Paolo and {Grefenstette}, Brian W. and {Kim}, Yunjin and {Kitaguchi}, Takao and {Koglin}, Jason E. and {Madsen}, Kristin K. and {Mao}, Peter H. and {Miyasaka}, Hiromasa and {Mori}, Kaya and {Perri}, Matteo and {Pivovaroff}, Michael J. and {Puccetti}, Simonetta and {Rana}, Vikram R. and {Westergaard}, Niels J. and {Willis}, Jason and {Zoglauer}, Andreas and {An}, Hongjun and {Bachetti}, Matteo and {Barri{\`e}re}, Nicolas M. and {Bellm}, Eric C. and {Bhalerao}, Varun and {Brejnholt}, Nicolai F. and {Fuerst}, Felix and {Liebe}, Carl C. and {Markwardt}, Craig B. and {Nynka}, Melania and {Vogel}, Julia K. and {Walton}, Dominic J. and {Wik}, Daniel R. and {Alexander}, David M. and {Cominsky}, Lynn R. and {Hornschemeier}, Ann E. and {Hornstrup}, Allan and {Kaspi}, Victoria M. and {Madejski}, Greg M. and {Matt}, Giorgio and {Molendi}, Silvano and {Smith}, David M. and {Tomsick}, John A. and {Ajello}, Marco and {Ballantyne}, David R. and {Balokovi{\'c}}, Mislav and {Barret}, Didier and {Bauer}, Franz E. and {Blandford}, Roger D. and {Brandt}, W. Niel and {Brenneman}, Laura W. and {Chiang}, James and {Chakrabarty}, Deepto and {Chenevez}, Jerome and {Comastri}, Andrea and {Dufour}, Francois and {Elvis}, Martin and {Fabian}, Andrew C. and {Farrah}, Duncan and {Fryer}, Chris L. and {Gotthelf}, Eric V. and {Grindlay}, Jonathan E. and {Helfand}, David J. and {Krivonos}, Roman and {Meier}, David L. and {Miller}, Jon M. and {Natalucci}, Lorenzo and {Ogle}, Patrick and {Ofek}, Eran O. and {Ptak}, Andrew and {Reynolds}, Stephen P. and {Rigby}, Jane R. and {Tagliaferri}, Gianpiero and {Thorsett}, Stephen E. and {Treister}, Ezequiel and {Urry}, C. Megan},
        title = "{The Nuclear Spectroscopic Telescope Array (NuSTAR) High-energy X-Ray Mission}",
      journal = {\apj},
         year = 2013,
        month = jun,
       volume = {770},
       number = {2},
          eid = {103},
        pages = {103},
          doi = {10.1088/0004-637X/770/2/103},
archivePrefix = {arXiv},
       eprint = {1301.7307},
 primaryClass = {astro-ph.IM},
       adsurl = {https://ui.adsabs.harvard.edu/abs/2013ApJ...770..103H}
}

@ARTICLE{Hutchings1983,
       author = {{Hutchings}, J.~B. and {Crampton}, D. and {Cowley}, A.~P.},
        title = "{A spectrographic orbit for LMC X-1 : another massive X-ray source ?}",
      journal = {\apjl},
         year = 1983,
        month = dec,
       volume = {275},
        pages = {L43-L47},
          doi = {10.1086/184168},
       adsurl = {https://ui.adsabs.harvard.edu/abs/1983ApJ...275L..43H}
}

@ARTICLE{Jana2021,
       author = {{Jana}, Arghajit and {Naik}, Sachindra and {Chatterjee}, Debjit and {Jaisawal}, Gaurava K.},
        title = "{NuSTAR and Swift observations of the extragalactic black hole X-ray binaries}",
      journal = {\mnras},
         year = 2021,
        month = nov,
       volume = {507},
       number = {4},
        pages = {4779-4787},
          doi = {10.1093/mnras/stab2448},
archivePrefix = {arXiv},
       eprint = {2108.10546},
 primaryClass = {astro-ph.HE},
       adsurl = {https://ui.adsabs.harvard.edu/abs/2021MNRAS.507.4779J}
}

@INPROCEEDINGS{Joye2003,
       author = {{Joye}, W.~A. and {Mandel}, E.},
        title = "{New Features of SAOImage DS9}",
    booktitle = {Astronomical Data Analysis Software and Systems XII},
         year = 2003,
       editor = {{Payne}, H.~E. and {Jedrzejewski}, R.~I. and {Hook}, R.~N.},
       series = {Astronomical Society of the Pacific Conference Series},
       volume = {295},
        month = jan,
        pages = {489},
       adsurl = {https://ui.adsabs.harvard.edu/abs/2003ASPC..295..489J}
}

@ARTICLE{Kaastra2016,
       author = {{Kaastra}, J.~S. and {Bleeker}, J.~A.~M.},
        title = "{Optimal binning of X-ray spectra and response matrix design}",
      journal = {\aap},
         year = 2016,
        month = mar,
       volume = {587},
          eid = {A151},
        pages = {A151},
          doi = {10.1051/0004-6361/201527395},
archivePrefix = {arXiv},
       eprint = {1601.05309},
 primaryClass = {astro-ph.IM},
       adsurl = {https://ui.adsabs.harvard.edu/abs/2016A&A...587A.151K}
}

@ARTICLE{Kravtsov2025,
       author = {{Kravtsov}, Vadim and {Bocharova}, Anastasiia and {Veledina}, Alexandra and {Poutanen}, Juri and {Hughes}, Andrew K. and {Dov{\v{c}}iak}, Michal and {Egron}, Elise and {Muleri}, Fabio and {Podgorny}, Jakub and {Svoboda}, Ji{\v{r}}i and {Forsblom}, Sofia V. and {Berdyugin}, Andrei V. and {Blinov}, Dmitry and {Bright}, Joe S. and {Carotenuto}, Francesco and {Green}, David A. and {Ingram}, Adam and {Liodakis}, Ioannis and {Mandarakas}, Nikos and {Nitindala}, Anagha P. and {Rhodes}, Lauren and {Trushkin}, Sergei A. and {Tsygankov}, Sergey S. and {Brigitte}, Ma{\"\i}mouna and {Di Marco}, Alessandro and {Iacolina}, Noemi and {Krawczynski}, Henric and {La Monaca}, Fabio and {Loktev}, Vladislav and {Mastroserio}, Guglielmo and {Petrucci}, Pierre-Olivier and {Pilia}, Maura and {Tombesi}, Francesco and {Zdziarski}, Andrzej A.},
        title = "{Variability of X-ray polarization of Cyg X-1}",
      journal = {\aap},
         year = 2025,
        month = sep,
       volume = {701},
          eid = {A115},
        pages = {A115},
          doi = {10.1051/0004-6361/202555411},
archivePrefix = {arXiv},
       eprint = {2505.03942},
 primaryClass = {astro-ph.HE},
       adsurl = {https://ui.adsabs.harvard.edu/abs/2025A&A...701A.115K}
}

@ARTICLE{Krawczynski2022,
       author = {{Krawczynski}, Henric and {Muleri}, Fabio and {Dov{\v{c}}iak}, Michal and {Veledina}, Alexandra and {Rodriguez Cavero}, Nicole and {Svoboda}, Jiri and {Ingram}, Adam and {Matt}, Giorgio and {Garcia}, Javier A. and {Loktev}, Vladislav and {Negro}, Michela and {Poutanen}, Juri and {Kitaguchi}, Takao and {Podgorn{\'y}}, Jakub and {Rankin}, John and {Zhang}, Wenda and {Berdyugin}, Andrei and {Berdyugina}, Svetlana V. and {Bianchi}, Stefano and {Blinov}, Dmitry and {Capitanio}, Fiamma and {Di Lalla}, Niccol{\`o} and {Draghis}, Paul and {Fabiani}, Sergio and {Kagitani}, Masato and {Kravtsov}, Vadim and {Kiehlmann}, Sebastian and {Latronico}, Luca and {Lutovinov}, Alexander A. and {Mandarakas}, Nikos and {Marin}, Fr{\'e}d{\'e}ric and {Marinucci}, Andrea and {Miller}, Jon M. and {Mizuno}, Tsunefumi and {Molkov}, Sergey V. and {Omodei}, Nicola and {Petrucci}, Pierre-Olivier and {Ratheesh}, Ajay and {Sakanoi}, Takeshi and {Semena}, Andrei N. and {Skalidis}, Raphael and {Soffitta}, Paolo and {Tennant}, Allyn F. and {Thalhammer}, Phillipp and {Tombesi}, Francesco and {Weisskopf}, Martin C. and {Wilms}, Joern and {Zhang}, Sixuan and {Agudo}, Iv{\'a}n and {Antonelli}, Lucio A. and {Bachetti}, Matteo and {Baldini}, Luca and {Baumgartner}, Wayne H. and {Bellazzini}, Ronaldo and {Bongiorno}, Stephen D. and {Bonino}, Raffaella and {Brez}, Alessandro and {Bucciantini}, Niccol{\`o} and {Castellano}, Simone and {Cavazzuti}, Elisabetta and {Ciprini}, Stefano and {Costa}, Enrico and {De Rosa}, Alessandra and {Del Monte}, Ettore and {Di Gesu}, Laura and {Di Marco}, Alessandro and {Donnarumma}, Immacolata and {Doroshenko}, Victor and {Ehlert}, Steven R. and {Enoto}, Teruaki and {Evangelista}, Yuri and {Ferrazzoli}, Riccardo and {Gunji}, Shuichi and {Hayashida}, Kiyoshi and {Heyl}, Jeremy and {Iwakiri}, Wataru and {Jorstad}, Svetlana G. and {Karas}, Vladimir and {Kolodziejczak}, Jeffery J. and {La Monaca}, Fabio and {Liodakis}, Ioannis and {Maldera}, Simone and {Manfreda}, Alberto and {Marscher}, Alan P. and {Marshall}, Herman L. and {Mitsuishi}, Ikuyuki and {Ng}, Chi-Yung and {O{\textquoteright}Dell}, Stephen L. and {Oppedisano}, Chiara and {Papitto}, Alessandro and {Pavlov}, George G. and {Peirson}, Abel L. and {Perri}, Matteo and {Pesce-Rollins}, Melissa and {Pilia}, Maura and {Possenti}, Andrea and {Puccetti}, Simonetta and {Ramsey}, Brian D. and {Romani}, Roger W. and {Sgr{\`o}}, Carmelo and {Slane}, Patrick and {Spandre}, Gloria and {Tamagawa}, Toru and {Tavecchio}, Fabrizio and {Taverna}, Roberto and {Tawara}, Yuzuru and {Thomas}, Nicholas E. and {Trois}, Alessio and {Tsygankov}, Sergey and {Turolla}, Roberto and {Vink}, Jacco and {Wu}, Kinwah and {Xie}, Fei and {Zane}, Silvia},
        title = "{Polarized x-rays constrain the disk-jet geometry in the black hole x-ray binary Cygnus X-1}",
      journal = {Science},
         year = 2022,
        month = nov,
       volume = {378},
       number = {6620},
        pages = {650-654},
          doi = {10.1126/science.add5399},
archivePrefix = {arXiv},
       eprint = {2206.09972},
 primaryClass = {astro-ph.HE},
       adsurl = {https://ui.adsabs.harvard.edu/abs/2022Sci...378..650K}
}

@ARTICLE{Krawczynski2025,
       author = {{Krawczynski}, Henric and {Hu}, Kun},
        title = "{The Cygnus X-1 Puzzle: Implications of X-Ray Polarization Measurements in the Soft and Hard States on the Properties of the Accretion Flow and the Emission Mechanisms}",
      journal = {\apj},
         year = 2025,
        month = nov,
       volume = {993},
       number = {1},
          eid = {54},
        pages = {54},
          doi = {10.3847/1538-4357/adfddf},
archivePrefix = {arXiv},
       eprint = {2506.01184},
 primaryClass = {astro-ph.HE},
       adsurl = {https://ui.adsabs.harvard.edu/abs/2025ApJ...993...54K}
}

@ARTICLE{Ling2026,
       author = {{Ling}, Yu-Shan and {Xie}, Fei and {Di Marco}, Alessandro and {La Monaca}, Fabio and {Ge}, Ming-Yu},
        title = "{Unchanged X-Ray Polarization during Accretion Dips in the Low Hard State of Cygnus X-1}",
      journal = {\apj},
         year = 2026,
        month = nov,
       volume = {1011},
       number = {2},
          eid = {198},
        pages = {198},
          doi = {10.3847/1538-4357/aea6d4},
archivePrefix = {arXiv},
       eprint = {2607.26144},
 primaryClass = {astro-ph.HE},
       adsurl = {https://ui.adsabs.harvard.edu/abs/2026ApJ..1011..198L}
}

@ARTICLE{Loktev22,
       author = {{Loktev}, Vladislav and {Veledina}, Alexandra and {Poutanen}, Juri},
        title = "{Analytical techniques for polarimetric imaging of accretion flows in the Schwarzschild metric}",
      journal = {\aap},
         year = 2022,
        month = apr,
       volume = {660},
          eid = {A25},
        pages = {A25},
          doi = {10.1051/0004-6361/202142360},
archivePrefix = {arXiv},
       eprint = {2109.04827},
 primaryClass = {astro-ph.HE},
       adsurl = {https://ui.adsabs.harvard.edu/abs/2022A&A...660A..25L}
}

@ARTICLE{Mark1969,
       author = {{Mark}, Hans and {Price}, R. and {Rodrigues}, R. and {Seward}, F.~D. and {Swift}, C.~D.},
        title = "{Detection of X-Rays from the Large Magellanic Cloud}",
      journal = {\apjl},
         year = 1969,
        month = mar,
       volume = {155},
        pages = {L143},
          doi = {10.1086/180322},
       adsurl = {https://ui.adsabs.harvard.edu/abs/1969ApJ...155L.143M}
}

@ARTICLE{Negueruela2002,
       author = {{Negueruela}, I. and {Coe}, M.~J.},
        title = "{The population of massive X-ray binaries. I. The Large Magellanic Cloud}",
      journal = {\aap},
         year = 2002,
        month = apr,
       volume = {385},
        pages = {517-532},
          doi = {10.1051/0004-6361:20020139},
archivePrefix = {arXiv},
       eprint = {astro-ph/0201451},
 primaryClass = {astro-ph},
       adsurl = {https://ui.adsabs.harvard.edu/abs/2002A&A...385..517N}
}

@ARTICLE{Nowak2001,
       author = {{Nowak}, M.~A. and {Wilms}, J. and {Heindl}, W.~A. and {Pottschmidt}, K. and {Dove}, J.~B. and {Begelman}, M.~C.},
        title = "{A good long look at the black hole candidates LMC X-1 and LMC X-3}",
      journal = {\mnras},
         year = 2001,
        month = jan,
       volume = {320},
       number = {3},
        pages = {316-326},
          doi = {10.1046/j.1365-8711.2001.03984.x},
archivePrefix = {arXiv},
       eprint = {astro-ph/0005487},
 primaryClass = {astro-ph},
       adsurl = {https://ui.adsabs.harvard.edu/abs/2001MNRAS.320..316N}
}

@ARTICLE{Orosz2009,
       author = {{Orosz}, Jerome A. and {Steeghs}, Danny and {McClintock}, Jeffrey E. and {Torres}, Manuel A.~P. and {Bochkov}, Ivan and {Gou}, Lijun and {Narayan}, Ramesh and {Blaschak}, Michael and {Levine}, Alan M. and {Remillard}, Ronald A. and {Bailyn}, Charles D. and {Dwyer}, Morgan M. and {Buxton}, Michelle},
        title = "{A New Dynamical Model for the Black Hole Binary LMC X-1}",
      journal = {\apj},
         year = 2009,
        month = may,
       volume = {697},
       number = {1},
        pages = {573-591},
          doi = {10.1088/0004-637X/697/1/573},
archivePrefix = {arXiv},
       eprint = {0810.3447},
 primaryClass = {astro-ph},
       adsurl = {https://ui.adsabs.harvard.edu/abs/2009ApJ...697..573O}
}

@ARTICLE{Pietrzynski2013,
       author = {{Pietrzy{\'n}ski}, G. and {Graczyk}, D. and {Gieren}, W. and {Thompson}, I.~B. and {Pilecki}, B. and {Udalski}, A. and {Soszy{\'n}ski}, I. and {Koz{\l}owski}, S. and {Konorski}, P. and {Suchomska}, K. and {Bono}, G. and {Moroni}, P.~G. Prada and {Villanova}, S. and {Nardetto}, N. and {Bresolin}, F. and {Kudritzki}, R.~P. and {Storm}, J. and {Gallenne}, A. and {Smolec}, R. and {Minniti}, D. and {Kubiak}, M. and {Szyma{\'n}ski}, M.~K. and {Poleski}, R. and {Wyrzykowski}, {\L}. and {Ulaczyk}, K. and {Pietrukowicz}, P. and {G{\'o}rski}, M. and {Karczmarek}, P.},
        title = "{An eclipsing-binary distance to the Large Magellanic Cloud accurate to two per cent}",
      journal = {\nat},
         year = 2013,
        month = mar,
       volume = {495},
       number = {7439},
        pages = {76-79},
          doi = {10.1038/nature11878},
archivePrefix = {arXiv},
       eprint = {1303.2063},
 primaryClass = {astro-ph.GA},
       adsurl = {https://ui.adsabs.harvard.edu/abs/2013Natur.495...76P}
}

@ARTICLE{Podgorny2023,
       author = {{Podgorn{\'y}}, J. and {Marra}, L. and {Muleri}, F. and {Rodriguez Cavero}, N. and {Ratheesh}, A. and {Dov{\v{c}}iak}, M. and {Miku{\v{s}}incov{\'a}}, R. and {Brigitte}, M. and {Steiner}, J.~F. and {Veledina}, A. and {Bianchi}, S. and {Krawczynski}, H. and {Svoboda}, J. and {Kaaret}, P. and {Matt}, G. and {Garc{\'\i}a}, J.~A. and {Petrucci}, P.-O. and {Lutovinov}, A. and {Semena}, A. and {Di Marco}, A. and {Negro}, M. and {Weisskopf}, M.~C. and {Ingram}, A. and {Poutanen}, J. and {Beheshtipour}, B. and {Chun}, S. and {Hu}, K. and {Mizuno}, T. and {Sixuan}, Z. and {Tombesi}, F. and {Zane}, S. and {Agudo}, I. and {Antonelli}, L.~A. and {Bachetti}, M. and {Baldini}, L. and {Baumgartner}, W.~H. and {Bellazzini}, R. and {Bongiorno}, S.~D. and {Bonino}, R. and {Brez}, A. and {Bucciantini}, N. and {Capitanio}, F. and {Castellano}, S. and {Cavazzuti}, E. and {Chen}, C. and {Ciprini}, S. and {Costa}, E. and {De Rosa}, A. and {Del Monte}, E. and {Di Gesu}, L. and {Di Lalla}, N. and {Donnarumma}, I. and {Doroshenko}, V. and {Ehlert}, S.~R. and {Enoto}, T. and {Evangelista}, Y. and {Fabiani}, S. and {Ferrazzoli}, R. and {Gunji}, S. and {Hayashida}, K. and {Heyl}, J. and {Iwakiri}, W. and {Jorstad}, S.~G. and {Karas}, V. and {Kislat}, F. and {Kitaguchi}, T. and {Kolodziejczak}, J.~J. and {La Monaca}, F. and {Latronico}, L. and {Liodakis}, I. and {Maldera}, S. and {Manfreda}, A. and {Marin}, F. and {Marinucci}, A. and {Marscher}, A.~P. and {Marshall}, H.~L. and {Massaro}, F. and {Mitsuishi}, I. and {Ng}, C.-Y. and {O'Dell}, S.~L. and {Omodei}, N. and {Oppedisano}, C. and {Papitto}, A. and {Pavlov}, G.~G. and {Peirson}, A.~L. and {Perri}, M. and {Pesce-Rollins}, M. and {Pilia}, M. and {Possenti}, A. and {Puccetti}, S. and {Ramsey}, B.~D. and {Rankin}, J. and {Roberts}, O.~J. and {Romani}, R.~W. and {Sgr{\`o}}, C. and {Slane}, P. and {Soffitta}, P. and {Spandre}, G. and {Swartz}, D.~A. and {Tamagawa}, T. and {Tavecchio}, F. and {Taverna}, R. and {Tawara}, Y. and {Tennant}, A.~F. and {Thomas}, N.~E. and {Trois}, A. and {Tsygankov}, S.~S. and {Turolla}, R. and {Vink}, J. and {Wu}, K. and {Xie}, F.},
        title = "{The first X-ray polarimetric observation of the black hole binary LMC X-1}",
      journal = {\mnras},
         year = 2023,
        month = dec,
       volume = {526},
       number = {4},
        pages = {5964-5975},
          doi = {10.1093/mnras/stad3103},
archivePrefix = {arXiv},
       eprint = {2303.12034},
 primaryClass = {astro-ph.HE},
       adsurl = {https://ui.adsabs.harvard.edu/abs/2023MNRAS.526.5964P}
}

@ARTICLE{Price1971,
       author = {{Price}, R.~E. and {Groves}, D.~J. and {Rodrigues}, R.~M. and {Seward}, F.~D. and {Swift}, C.~D. and {Toor}, A.},
        title = "{X-Rays from the Magellanic Clouds}",
      journal = {\apjl},
         year = 1971,
        month = aug,
       volume = {168},
        pages = {L7},
          doi = {10.1086/180773},
       adsurl = {https://ui.adsabs.harvard.edu/abs/1971ApJ...168L...7P}
}

@BOOK{Sobolev63,
       author = {{Sobolev}, Viktor Viktorovich},
        title = "{A treatise on radiative transfer.}",
         year = 1963,
       adsurl = {https://ui.adsabs.harvard.edu/abs/1963trt..book.....S}
}

@ARTICLE{Soffitta2021,
       author = {{Soffitta}, Paolo and {Baldini}, Luca and {Bellazzini}, Ronaldo and {Costa}, Enrico and {Latronico}, Luca and {Muleri}, Fabio and {Del Monte}, Ettore and {Fabiani}, Sergio and {Minuti}, Massimo and {Pinchera}, Michele and {Sgro'}, Carmelo and {Spandre}, Gloria and {Trois}, Alessio and {Amici}, Fabrizio and {Andersson}, Hans and {Attina'}, Primo and {Bachetti}, Matteo and {Barbanera}, Mattia and {Borotto}, Fabio and {Brez}, Alessandro and {Brienza}, Daniele and {Caporale}, Ciro and {Cardelli}, Claudia and {Carpentiero}, Rita and {Castellano}, Simone and {Castronuovo}, Marco and {Cavalli}, Luca and {Cavazzuti}, Elisabetta and {Ceccanti}, Marco and {Centrone}, Mauro and {Ciprini}, Stefano and {Citraro}, Saverio and {D'Amico}, Fabio and {D'Alba}, Elisa and {Di Cosimo}, Sergio and {Di Lalla}, Niccolo' and {Di Marco}, Alessandro and {Di Persio}, Giuseppe and {Donnarumma}, Immacolata and {Evangelista}, Yuri and {Ferrazzoli}, Riccardo and {Hayato}, Asami and {Kitaguchi}, Takao and {La Monaca}, Fabio and {Lefevre}, Carlo and {Loffredo}, Pasqualino and {Lorenzi}, Paolo and {Lucchesi}, Leonardo and {Magazzu}, Carlo and {Maldera}, Simone and {Manfreda}, Alberto and {Mangraviti}, Elio and {Marengo}, Marco and {Matt}, Giorgio and {Mereu}, Paolo and {Morbidini}, Alfredo and {Mosti}, Federico and {Nakano}, Toshio and {Nasimi}, Hikmat and {Negri}, Barbara and {Nenonen}, Seppo and {Nuti}, Alessio and {Orsini}, Leonardo and {Perri}, Matteo and {Pesce-Rollins}, Melissa and {Piazzolla}, Raffaele and {Pilia}, Maura and {Profeti}, Alessandro and {Puccetti}, Simonetta and {Rankin}, John and {Ratheesh}, Ajay and {Rubini}, Alda and {Santoli}, Francesco and {Sarra}, Paolo and {Scalise}, Emanuele and {Sciortino}, Andrea and {Tamagawa}, Toru and {Tardiola}, Marcello and {Tobia}, Antonino and {Vimercati}, Marco and {Xie}, Fei},
        title = "{The Instrument of the Imaging X-Ray Polarimetry Explorer}",
      journal = {\aj},
         year = 2021,
        month = nov,
       volume = {162},
       number = {5},
          eid = {208},
        pages = {208},
          doi = {10.3847/1538-3881/ac19b0},
archivePrefix = {arXiv},
       eprint = {2108.00284},
 primaryClass = {astro-ph.IM},
       adsurl = {https://ui.adsabs.harvard.edu/abs/2021AJ....162..208S}
}

@ARTICLE{Steiner2012,
       author = {{Steiner}, James F. and {Reis}, Rubens C. and {Fabian}, Andrew C. and {Remillard}, Ronald A. and {McClintock}, Jeffrey E. and {Gou}, Lijun and {Cooke}, Ryan and {Brenneman}, Laura W. and {Sanders}, Jeremy S.},
        title = "{A broad iron line in LMC X-1}",
      journal = {\mnras},
         year = 2012,
        month = dec,
       volume = {427},
       number = {3},
        pages = {2552-2561},
          doi = {10.1111/j.1365-2966.2012.22128.x},
archivePrefix = {arXiv},
       eprint = {1209.3269},
 primaryClass = {astro-ph.HE},
       adsurl = {https://ui.adsabs.harvard.edu/abs/2012MNRAS.427.2552S}
}

@ARTICLE{Steiner2024,
       author = {{Steiner}, James F. and {Nathan}, Edward and {Hu}, Kun and {Krawczynski}, Henric and {Dov{\v{c}}iak}, Michal and {Veledina}, Alexandra and {Muleri}, Fabio and {Svoboda}, Jiri and {Alabarta}, Kevin and {Parra}, Maxime and {Bhargava}, Yash and {Matt}, Giorgio and {Poutanen}, Juri and {Petrucci}, Pierre-Olivier and {Tennant}, Allyn F. and {Baglio}, M. Cristina and {Baldini}, Luca and {Barnier}, Samuel and {Bhattacharyya}, Sudip and {Bianchi}, Stefano and {Brigitte}, Maimouna and {Cabezas}, Mauricio and {Cangemi}, Floriane and {Capitanio}, Fiamma and {Casey}, Jacob and {Rodriguez Cavero}, Nicole and {Castellano}, Simone and {Cavazzuti}, Elisabetta and {Chun}, Sohee and {Churazov}, Eugene and {Costa}, Enrico and {Di Lalla}, Niccol{\`o} and {Di Marco}, Alessandro and {Egron}, Elise and {Ewing}, Melissa and {Fabiani}, Sergio and {Garc{\'\i}a}, Javier A. and {Green}, David A. and {Grinberg}, Victoria and {Hadrava}, Petr and {Ingram}, Adam and {Kaaret}, Philip and {Kislat}, Fabian and {Kitaguchi}, Takao and {Kravtsov}, Vadim and {Kub{\'a}tov{\'a}}, Brankica and {La Monaca}, Fabio and {Latronico}, Luca and {Loktev}, Vladislav and {Malacaria}, Christian and {Marin}, Fr{\'e}d{\'e}ric and {Marinucci}, Andrea and {Maryeva}, Olga and {Mastroserio}, Guglielmo and {Mizuno}, Tsunefumi and {Negro}, Michela and {Omodei}, Nicola and {Podgorn{\'y}}, Jakub and {Rankin}, John and {Ratheesh}, Ajay and {Rhodes}, Lauren and {Russell}, David M. and {{\v{S}}lechta}, Miroslav and {Soffitta}, Paolo and {Spooner}, Sean and {Suleimanov}, Valery and {Tombesi}, Francesco and {Trushkin}, Sergei A. and {Weisskopf}, Martin C. and {Zane}, Silvia and {Zdziarski}, Andrzej A. and {Zhang}, Sixuan and {Zhang}, Wenda and {Zhou}, Menglei and {Agudo}, Iv{\'a}n and {Antonelli}, Lucio A. and {Bachetti}, Matteo and {Baumgartner}, Wayne H. and {Bellazzini}, Ronaldo and {Bongiorno}, Stephen D. and {Bonino}, Raffaella and {Brez}, Alessandro and {Bucciantini}, Niccol{\`o} and {Chen}, Chien-Ting and {Ciprini}, Stefano and {De Rosa}, Alessandra and {Del Monte}, Ettore and {Di Gesu}, Laura and {Donnarumma}, Immacolata and {Doroshenko}, Victor and {Ehlert}, Steven R. and {Enoto}, Teruaki and {Evangelista}, Yuri and {Ferrazzoli}, Riccardo and {Gunji}, Shuichi and {Hayashida}, Kiyoshi and {Heyl}, Jeremy and {Iwakiri}, Wataru and {Jorstad}, Svetlana G. and {Karas}, Vladimir and {Kolodziejczak}, Jeffery J. and {Liodakis}, Ioannis and {Maldera}, Simone and {Manfreda}, Alberto and {Marscher}, Alan P. and {Marshall}, Herman L. and {Massaro}, Francesco and {Mitsuishi}, Ikuyuki and {Ng}, Chi-Yung and {O'Dell}, Stephen L. and {Oppedisano}, Chiara and {Papitto}, Alessandro and {Pavlov}, George G. and {Peirson}, Abel L. and {Perri}, Matteo and {Pesce-Rollins}, Melissa and {Pilia}, Maura and {Possenti}, Andrea and {Puccetti}, Simonetta and {Ramsey}, Brian D. and {Roberts}, Oliver J. and {Romani}, Roger W. and {Sgr{\`o}}, Carmelo and {Slane}, Patrick and {Spandre}, Gloria and {Swartz}, Douglas A. and {Tamagawa}, Toru and {Tavecchio}, Fabrizio and {Taverna}, Roberto and {Tawara}, Yuzuru and {Thomas}, Nicholas E. and {Trois}, Alessio and {Tsygankov}, Sergey S. and {Turolla}, Roberto and {Vink}, Jacco and {Wu}, Kinwah and {Xie}, Fei},
        title = "{An IXPE-led X-Ray Spectropolarimetric Campaign on the Soft State of Cygnus X-1: X-Ray Polarimetric Evidence for Strong Gravitational Lensing}",
      journal = {\apjl},
         year = 2024,
        month = jul,
       volume = {969},
       number = {2},
          eid = {L30},
        pages = {L30},
          doi = {10.3847/2041-8213/ad58e4},
archivePrefix = {arXiv},
       eprint = {2406.12014},
 primaryClass = {astro-ph.HE},
       adsurl = {https://ui.adsabs.harvard.edu/abs/2024ApJ...969L..30S}
}

@ARTICLE{Veledina24,
       author = {{Veledina}, Alexandra and {Muleri}, Fabio and {Poutanen}, Juri and {Podgorn{\'y}}, Jakub and {Dov{\v{c}}iak}, Michal and {Capitanio}, Fiamma and {Churazov}, Eugene and {De Rosa}, Alessandra and {Di Marco}, Alessandro and {Forsblom}, Sofia V. and {Kaaret}, Philip and {Krawczynski}, Henric and {La Monaca}, Fabio and {Loktev}, Vladislav and {Lutovinov}, Alexander A. and {Molkov}, Sergey V. and {Mushtukov}, Alexander A. and {Ratheesh}, Ajay and {Rodriguez Cavero}, Nicole and {Steiner}, James F. and {Sunyaev}, Rashid A. and {Tsygankov}, Sergey S. and {Weisskopf}, Martin C. and {Zdziarski}, Andrzej A. and {Bianchi}, Stefano and {Bright}, Joe S. and {Bursov}, Nikolaj and {Costa}, Enrico and {Egron}, Elise and {Garcia}, Javier A. and {Green}, David A. and {Gurwell}, Mark and {Ingram}, Adam and {Kajava}, Jari J.~E. and {Kale}, Ruta and {Kraus}, Alex and {Malyshev}, Denys and {Marin}, Fr{\'e}d{\'e}ric and {Matt}, Giorgio and {McCollough}, Michael and {Mereminskiy}, Ilya A. and {Nizhelsky}, Nikolaj and {Piano}, Giovanni and {Pilia}, Maura and {Pittori}, Carlotta and {Rao}, Ramprasad and {Righini}, Simona and {Soffitta}, Paolo and {Shevchenko}, Anton and {Svoboda}, Jiri and {Tombesi}, Francesco and {Trushkin}, Sergei A. and {Tsybulev}, Peter and {Ursini}, Francesco and {Wu}, Kinwah and {Agudo}, Iv{\'a}n and {Antonelli}, Lucio A. and {Bachetti}, Matteo and {Baldini}, Luca and {Baumgartner}, Wayne H. and {Bellazzini}, Ronaldo and {Bongiorno}, Stephen D. and {Bonino}, Raffaella and {Brez}, Alessandro and {Bucciantini}, Niccol{\`o} and {Castellano}, Simone and {Cavazzuti}, Elisabetta and {Chen}, Chien-Ting and {Ciprini}, Stefano and {Del Monte}, Ettore and {Di Gesu}, Laura and {Di Lalla}, Niccol{\`o} and {Donnarumma}, Immacolata and {Doroshenko}, Victor and {Ehlert}, Steven R. and {Enoto}, Teruaki and {Evangelista}, Yuri and {Fabiani}, Sergio and {Ferrazzoli}, Riccardo and {Gunji}, Shuichi and {Hayashida}, Kiyoshi and {Heyl}, Jeremy and {Iwakiri}, Wataru and {Jorstad}, Svetlana G. and {Karas}, Vladimir and {Kislat}, Fabian and {Kitaguchi}, Takao and {Kolodziejczak}, Jeffery J. and {Latronico}, Luca and {Liodakis}, Ioannis and {Maldera}, Simone and {Manfreda}, Alberto and {Marinucci}, Andrea and {Marscher}, Alan P. and {Marshall}, Herman L. and {Massaro}, Francesco and {Mitsuishi}, Ikuyuki and {Mizuno}, Tsunefumi and {Negro}, Michela and {Ng}, Chi-Yung and {O'Dell}, Stephen L. and {Omodei}, Nicola and {Oppedisano}, Chiara and {Papitto}, Alessandro and {Pavlov}, George G. and {Peirson}, Abel L. and {Perri}, Matteo and {Pesce-Rollins}, Melissa and {Petrucci}, Pierre-Olivier and {Possenti}, Andrea and {Puccetti}, Simonetta and {Ramsey}, Brian D. and {Rankin}, John and {Roberts}, Oliver and {Romani}, Roger W. and {Sgr{\`o}}, Carmelo and {Slane}, Patrick and {Spandre}, Gloria and {Swartz}, Doug and {Tamagawa}, Toru and {Tavecchio}, Fabrizio and {Taverna}, Roberto and {Tawara}, Yuzuru and {Tennant}, Allyn F. and {Thomas}, Nicholas E. and {Trois}, Alessio and {Turolla}, Roberto and {Vink}, Jacco and {Xie}, Fei and {Zane}, Silvia},
        title = "{Cygnus X-3 revealed as a Galactic ultraluminous X-ray source by IXPE}",
      journal = {Nature Astronomy},
         year = 2024,
        month = aug,
       volume = {8},
        pages = {1031-1046},
          doi = {10.1038/s41550-024-02294-9},
archivePrefix = {arXiv},
       eprint = {2303.01174},
 primaryClass = {astro-ph.HE},
       adsurl = {https://ui.adsabs.harvard.edu/abs/2024NatAs...8.1031V}
}

@ARTICLE{Vern95,
       author = {{Verner}, D.~A. and {Yakovlev}, D.~G.},
        title = "{Analytic FITS for partial photoionization cross sections.}",
      journal = {\aaps},
         year = 1995,
        month = jan,
       volume = {109},
        pages = {125-133},
       adsurl = {https://ui.adsabs.harvard.edu/abs/1995A&AS..109..125V}
}

@ARTICLE{Weisskopf2022,
       author = {{Weisskopf}, Martin C. and {Soffitta}, Paolo and {Baldini}, Luca and {Ramsey}, Brian D. and {O'Dell}, Stephen L. and {Romani}, Roger W. and {Matt}, Giorgio and {Deininger}, William D. and {Baumgartner}, Wayne H. and {Bellazzini}, Ronaldo and {Costa}, Enrico and {Kolodziejczak}, Jeffery J. and {Latronico}, Luca and {Marshall}, Herman L. and {Muleri}, Fabio and {Bongiorno}, Stephen D. and {Tennant}, Allyn and {Bucciantini}, Niccolo and {Dovciak}, Michal and {Marin}, Frederic and {Marscher}, Alan and {Poutanen}, Juri and {Slane}, Pat and {Turolla}, Roberto and {Kalinowski}, William and {Di Marco}, Alessandro and {Fabiani}, Sergio and {Minuti}, Massimo and {La Monaca}, Fabio and {Pinchera}, Michele and {Rankin}, John and {Sgro'}, Carmelo and {Trois}, Alessio and {Xie}, Fei and {Alexander}, Cheryl and {Allen}, D. Zachery and {Amici}, Fabrizio and {Andersen}, Jason and {Antonelli}, Angelo and {Antoniak}, Spencer and {Attin{\`a}}, Primo and {Barbanera}, Mattia and {Bachetti}, Matteo and {Baggett}, Randy M. and {Bladt}, Jeff and {Brez}, Alessandro and {Bonino}, Raffaella and {Boree}, Christopher and {Borotto}, Fabio and {Breeding}, Shawn and {Brienza}, Daniele and {Bygott}, H. Kyle and {Caporale}, Ciro and {Cardelli}, Claudia and {Carpentiero}, Rita and {Castellano}, Simone and {Castronuovo}, Marco and {Cavalli}, Luca and {Cavazzuti}, Elisabetta and {Ceccanti}, Marco and {Centrone}, Mauro and {Citraro}, Saverio and {D'Amico}, Fabio and {D'Alba}, Elisa and {Di Gesu}, Laura and {Del Monte}, Ettore and {Dietz}, Kurtis L. and {Di Lalla}, Niccolo' and {Persio}, Giuseppe Di and {Dolan}, David and {Donnarumma}, Immacolata and {Evangelista}, Yuri and {Ferrant}, Kevin and {Ferrazzoli}, Riccardo and {Ferrie}, MacKenzie and {Footdale}, Joseph and {Forsyth}, Brent and {Foster}, Michelle and {Garelick}, Benjamin and {Gunji}, Shuichi and {Gurnee}, Eli and {Head}, Michael and {Hibbard}, Grant and {Johnson}, Samantha and {Kelly}, Erik and {Kilaru}, Kiranmayee and {Lefevre}, Carlo and {Roy}, Shelley Le and {Loffredo}, Pasqualino and {Lorenzi}, Paolo and {Lucchesi}, Leonardo and {Maddox}, Tyler and {Magazzu}, Guido and {Maldera}, Simone and {Manfreda}, Alberto and {Mangraviti}, Elio and {Marengo}, Marco and {Marrocchesi}, Alessandra and {Massaro}, Francesco and {Mauger}, David and {McCracken}, Jeffrey and {McEachen}, Michael and {Mize}, Rondal and {Mereu}, Paolo and {Mitchell}, Scott and {Mitsuishi}, Ikuyuki and {Morbidini}, Alfredo and {Mosti}, Federico and {Nasimi}, Hikmat and {Negri}, Barbara and {Negro}, Michela and {Nguyen}, Toan and {Nitschke}, Isaac and {Nuti}, Alessio and {Onizuka}, Mitch and {Oppedisano}, Chiara and {Orsini}, Leonardo and {Osborne}, Darren and {Pacheco}, Richard and {Paggi}, Alessandro and {Painter}, Will and {Pavelitz}, Steven D. and {Pentz}, Christina and {Piazzolla}, Raffaele and {Perri}, Matteo and {Pesce-Rollins}, Melissa and {Peterson}, Colin and {Pilia}, Maura and {Profeti}, Alessandro and {Puccetti}, Simonetta and {Ranganathan}, Jaganathan and {Ratheesh}, Ajay and {Reedy}, Lee and {Root}, Noah and {Rubini}, Alda and {Ruswick}, Stephanie and {Sanchez}, Javier and {Sarra}, Paolo and {Santoli}, Francesco and {Scalise}, Emanuele and {Sciortino}, Andrea and {Schroeder}, Christopher and {Seek}, Tim and {Sosdian}, Kalie and {Spandre}, Gloria and {Speegle}, Chet O. and {Tamagawa}, Toru and {Tardiola}, Marcello and {Tobia}, Antonino and {Thomas}, Nicholas E. and {Valerie}, Robert and {Vimercati}, Marco and {Walden}, Amy L. and {Weddendorf}, Bruce and {Wedmore}, Jeffrey and {Welch}, David and {Zanetti}, Davide and {Zanetti}, Francesco},
        title = "{The Imaging X-Ray Polarimetry Explorer (IXPE): Pre-Launch}",
      journal = {Journal of Astronomical Telescopes, Instruments, and Systems},
         year = 2022,
        month = apr,
       volume = {8},
       number = {2},
          eid = {026002},
        pages = {026002},
          doi = {10.1117/1.JATIS.8.2.026002},
archivePrefix = {arXiv},
       eprint = {2112.01269},
 primaryClass = {astro-ph.IM},
       adsurl = {https://ui.adsabs.harvard.edu/abs/2022JATIS...8b6002W}
}

@ARTICLE{Wilms2000,
       author = {{Wilms}, J. and {Allen}, A. and {McCray}, R.},
        title = "{On the Absorption of X-Rays in the Interstellar Medium}",
      journal = {\apj},
         year = 2000,
        month = oct,
       volume = {542},
       number = {2},
        pages = {914-924},
          doi = {10.1086/317016},
archivePrefix = {arXiv},
       eprint = {astro-ph/0008425},
 primaryClass = {astro-ph},
       adsurl = {https://ui.adsabs.harvard.edu/abs/2000ApJ...542..914W}
}

@ARTICLE{Zdziarski2024,
       author = {{Zdziarski}, Andrzej A. and {Banerjee}, Srimanta and {Chand}, Swadesh and {Dewangan}, Gulab and {Misra}, Ranjeev and {Szanecki}, Micha{\l} and {Nied{\'z}wiecki}, Andrzej},
        title = "{Black Hole Spin Measurements in LMC X-1 and Cyg X-1 Are Highly Model Dependent}",
      journal = {\apj},
         year = 2024,
        month = feb,
       volume = {962},
       number = {2},
          eid = {101},
        pages = {101},
          doi = {10.3847/1538-4357/ad1b60},
archivePrefix = {arXiv},
       eprint = {2308.06167},
 primaryClass = {astro-ph.HE},
       adsurl = {https://ui.adsabs.harvard.edu/abs/2024ApJ...962..101Z}
}

@ARTICLE{LaMonaca25,
       author = {{La Monaca}, Fabio and {Di Marco}, Alessandro and {Coti Zelati}, Francesco and {Bobrikova}, Anna and {Ludlam}, Renee M. and {Poutanen}, Juri and {Marino}, Alessio and {Li}, Songwei and {Xie}, Fei and {Feng}, Hua and et al.},
        title = "{X-ray spectropolarimetric characterisation of the Z source GX 340+0 in the normal branch}",
      journal = {\aap},
         year = 2025,
        month = oct,
       volume = {702},
          eid = {A101},
        pages = {A101},
          doi = {10.1051/0004-6361/202555134},
archivePrefix = {arXiv},
       eprint = {2508.13278},
 primaryClass = {astro-ph.HE},
       adsurl = {https://ui.adsabs.harvard.edu/abs/2025A&A...702A.101L}
}

@ARTICLE{Zdziarski2026,
       author = {{Zdziarski}, Andrzej A. and {Marcel}, Gr{\'e}goire and {Veledina}, Alexandra and {Olejak}, Aleksandra and {Lan{\v{c}}ov{\'a}}, Debora},
        title = "{Spins of black holes in X-ray binaries and the tension with the gravitational wave measurements}",
      journal = {\nar},
         year = 2026,
        month = jun,
       volume = {102},
          eid = {101746},
        pages = {101746},
          doi = {10.1016/j.newar.2025.101746},
archivePrefix = {arXiv},
       eprint = {2506.00623},
 primaryClass = {astro-ph.HE},
       adsurl = {https://ui.adsabs.harvard.edu/abs/2026NewAR.10201746Z}
}

@ARTICLE{Done2007,
       author = {{Done}, Chris and {Gierli{\'n}ski}, Marek and {Kubota}, Aya},
        title = "{Modelling the behaviour of accretion flows in X-ray binaries. Everything you always wanted to know about accretion but were afraid to ask}",
      journal = {\aapr},
         year = 2007,
        month = dec,
       volume = {15},
       number = {1},
        pages = {1-66},
          doi = {10.1007/s00159-007-0006-1},
archivePrefix = {arXiv},
       eprint = {0708.0148},
 primaryClass = {astro-ph},
       adsurl = {https://ui.adsabs.harvard.edu/abs/2007A&ARv..15....1D}
}

@ARTICLE{Marcel2026,
       author = {{Marcel}, Gr{\'e}goire and {Tetarenko}, Bailey and {Ingram}, Adam and {Maccarone}, Tom and {Veledina}, Alexandra and {Charles}, Phil},
        title = "{Accreting Stellar-Mass Black Holes}",
      journal = {\ssr},
         year = 2026,
        month = oct,
       volume = {222},
       number = {8},
          eid = {80},
        pages = {80},
          doi = {10.1007/s11214-026-01331-5},
archivePrefix = {arXiv},
       eprint = {2606.19952},
 primaryClass = {astro-ph.HE},
       adsurl = {https://ui.adsabs.harvard.edu/abs/2026SSRv..222...80M}
}

\begin{appendix}
\section{Fitting of the Folded Stokes parameters} \label{app:Folded_QU}

\begin{figure}[!ht]
\centering
\includegraphics[width=\linewidth]{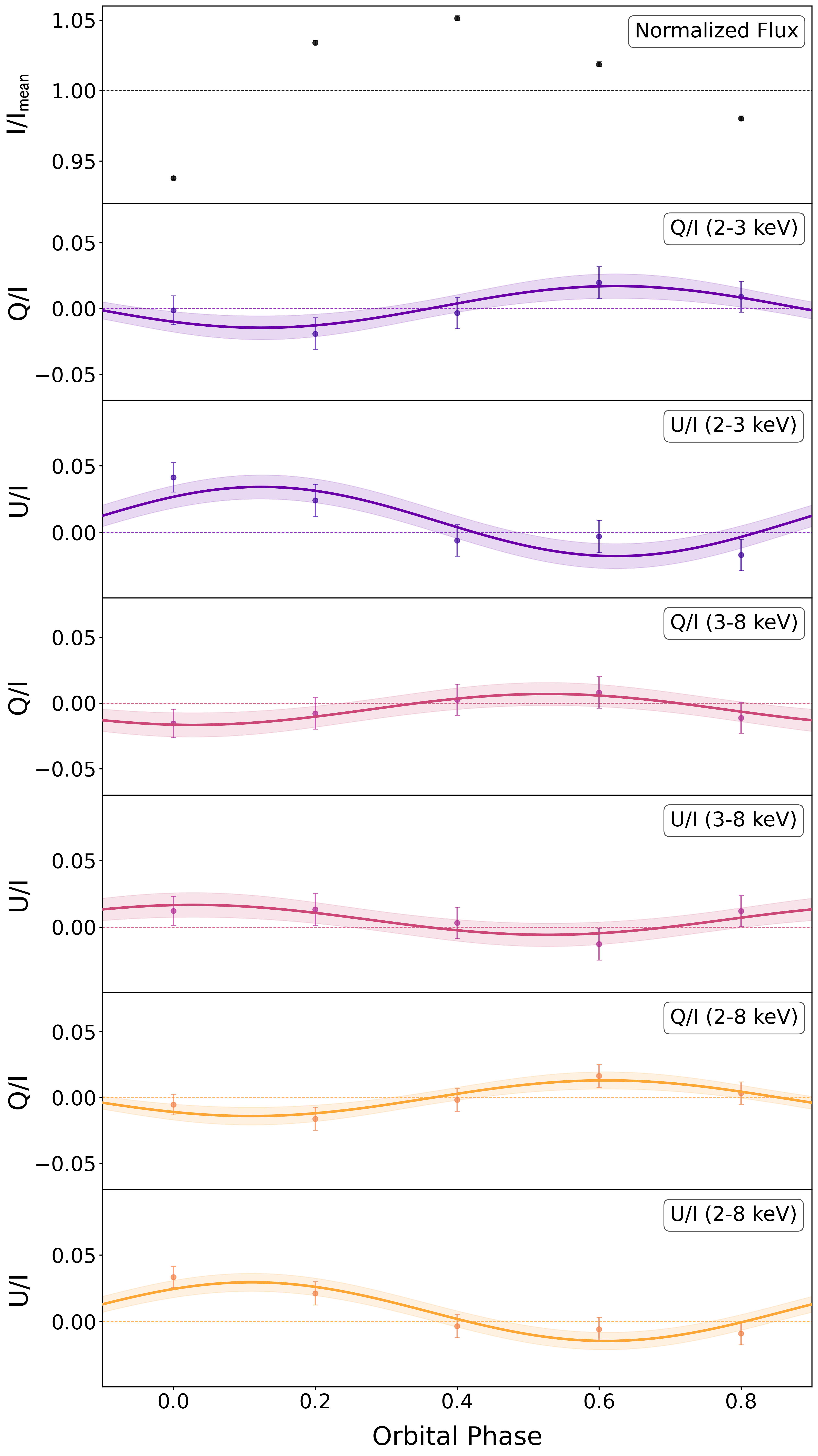}
\caption{The folded normalized light curve in the 2--8\,keV, Q/I, and U/I versus time in the 2--3\,keV, 3--8\,keV, and 2--8\,keV bands. Data points are extracted from time bins 0.2 orbital phase wide; errors are reported at 68\% CL. The red curves are cosine fits with the period fixed at 3.90917 days \citep{Orosz2009}.}
\label{fig:Folded_QU_vs_time}
\end{figure}

In Figure~\ref{fig:Folded_QU_vs_time}, we also folded $I$ in the 2--8\,keV, $Q/I$ and $U/I$ in the 2--3\,keV, 3--8\,keV, and 2--8\,keV bands, respectively, considering an orbital period fixed to 3.90917\,days \citep{Orosz2009}. The data were binned in intervals covering 0.2 orbital phases and fitted with the equations:
\begin{equation}\label{eq:Folded_QUcos}
\begin{aligned}
(Q/I)(\phi) &= A_Q \cos\left(2\pi(\phi + \phi_0)\right) + C_Q\\
(U/I)(\phi) &= A_U \cos\left(2\pi(\phi + \phi_0) + \pi\right) + C_U
\end{aligned}
\end{equation}

The light curve exhibits a flux decrease of 6.3\%, consistent with the reported modulation amplitude of ${\sim}7\%$ \citep{Podgorny2023}. The phase-merged $Q/I$ and $U/I$ values are listed in Table~\ref{tab:QU_phase_Parameters} and the best‑fit parameters in Table~\ref{tab:folded_cos_Parameters}. The $P_{\rm mod}$ obtained from the folded data are found to be in agreement with those derived in Table~\ref{tab:cos_Parameters}.

Under a phase-independent constant model, the best-fit levels are $C_{Q} = (0.1\pm0.5)\%$, $(-0.5\pm0.5)\%$ and $(-0.1\pm0.4)\%$, while $C_{U} = (0.9\pm0.5)\%$, $(0.6\pm0.5)\%$ and $(0.8\pm0.4)\%$ for 2--3\,keV, 3--8\,keV, and 2--8\,keV, respectively.

The constant model is disfavored at 2--3\,keV and 2--8\,keV, where $\chi^{2}/d.o.f. = 23/8$ and $28/8$, compared with $7/5$ and $5/5$ for the cosine model. In 3--8\,keV, the data are equally consistent with the constant model ($\chi^{2}/d.o.f. = 6/8$) and the cosine model ($\chi^{2}/d.o.f. = 1.3/5$); no significant modulation is required.

\begin{table*}
\centering
\caption{$Q/I$, $U/I$, PD and PA of every 0.2 phase in the 2--3\,keV, 3--8\,keV, and 2--8\,keV bands. Errors are reported at 68\%\,CL. Upper limits are reported at 90\%\,CL.}
\label{tab:QU_phase_Parameters}
\begin{tabular}{l c c c c c c}
\hline
\multicolumn{1}{c}{Phase} & \multicolumn{2}{c}{2--3\,keV} & \multicolumn{2}{c}{3--8\,keV} & \multicolumn{2}{c}{2--8\,keV} \\
\cline{2-7}
 & $Q/I$ (\%) & $U/I$ (\%) & $Q/I$ (\%) & $U/I$ (\%) & $Q/I$ (\%) & $U/I$ (\%) \\
\hline
-0.1--0.1 & $-0.1\pm1.1$ & $4.2\pm1.1$ & $-1.6\pm1.1$ & $1.2\pm1.1$ & $-0.5\pm0.8$ & $3.4\pm0.8$ \\
0.1--0.3 & $-1.9\pm1.2$ & $2.4\pm1.2$ & $-0.8\pm1.2$ & $1.3\pm1.2$ & $-1.6\pm0.9$ & $2.1\pm0.9$ \\
0.3--0.5 & $-0.3\pm1.2$ & $-0.6\pm1.2$ & $0.3\pm1.2$ & $0.3\pm1.2$ & $-0.2\pm0.9$ & $-0.4\pm0.9$ \\
0.5--0.7 & $2.0\pm1.2$ & $-0.3\pm1.2$ & $0.8\pm1.2$ & $-1.3\pm1.2$ & $1.7\pm0.9$ & $-0.6\pm0.9$ \\
0.7--0.9 & $0.9\pm1.2$ & $-1.7\pm1.2$ & $-1.1\pm1.2$ & $1.2\pm1.2$ & $0.3\pm0.9$ & $-0.9\pm0.9$ \\
\hline
 & $PD$ (\%) & $PA$ ($^\circ$) & $PD$ (\%) & $PA$ ($^\circ$) & $PD$ (\%) & $PA$ ($^\circ$) \\
\hline
-0.1--0.1 & $4.2\pm1.1$ & $46\pm8$ & $2.0\pm1.1$ & $71\pm16$ & $3.4\pm0.8$ & $50\pm7$ \\
0.1--0.3 & $3.1\pm1.2$ & $64\pm11$ & $1.5\pm1.2$ & $61\pm22$ & $2.7\pm0.9$ & $64\pm9$ \\
0.3--0.5 & $<2.2$ & $-$ & $<1.9$ & $-$ & $<1.5$ & $-$ \\
0.5--0.7 & $2.0\pm1.2$ & $-4\pm18$ & $1.5\pm1.2$ & $-29\pm23$ & $1.8\pm0.9$ & $-10\pm14$ \\
0.7--0.9 & $1.9\pm1.2$ & $-31\pm18$ & $1.7\pm1.2$ & $67\pm20$ & $1.0\pm0.9$ & $-35\pm25$ \\
\hline
\end{tabular}
\end{table*}

\begin{table}[!htb]
\centering
\caption{Best-fit parameters of the cosine model for the folded $Q/I$ and $U/I$ Stokes parameters in the 2--3\,keV, 3--8\,keV, and 2--8\,keV bands. Errors are reported at 68\%\,CL.}
\label{tab:folded_cos_Parameters}
\begin{tabular}{lcc}
\hline
Energy & Parameter & Value \\
\hline
2--3\,keV & $A_{Q}$ & $0.016\pm0.008$ \\
 & $A_{U}$ & $0.026\pm0.008$ \\
 & $\phi_0$ & $2.4\pm0.2$ \\
 & $C_{Q}$ & $0.001\pm0.006$ \\
 & $C_{U}$ & $0.008\pm0.006$ \\
 & $\chi^{2}/d.o.f.$ & 7/5 \\
\hline
3--8\,keV & $A_{Q}$ & $0.012\pm0.007$ \\
 & $A_{U}$ & $0.011\pm0.007$ \\
 & $\phi_0$ & $3.0\pm0.5$ \\
 & $C_{Q}$ & $-0.005\pm0.005$ \\
 & $C_{U}$ & $0.005\pm0.005$ \\
 & $\chi^{2}/d.o.f.$ & 1.3/5 \\
\hline
2--8\,keV & $A_{Q}$ & $0.014\pm0.005$ \\
 & $A_{U}$ & $0.022\pm0.005$ \\
 & $\phi_0$ & $2.5\pm0.2$ \\
 & $C_{Q}$ & $-0.001\pm0.004$ \\
 & $C_{U}$ & $0.008\pm0.004$ \\
 & $\chi^{2}/d.o.f.$ & 5/5 \\
\hline
\end{tabular}
\end{table}

\section{MBPO Model and Parameters} \label{app:mbpo}

To account for cross-calibration uncertainties among the different instruments, we multiplied the spectral model for each dataset by the \texttt{MBPO} multiplicative model, following the approach used in \citealt{Krawczynski2022} and \citealt{Ling2026}.
The \texttt{MBPO} model consists of a broken power-law function that adjusts the effective area ratios between instruments. Specifically, it modifies the continuum slope at low and high energies with two photon-index offsets ($d\Gamma_{1}$ and $d\Gamma_{2}$) and a break energy ($E_{br}$), together with a normalization constant (\texttt{norm}).

For the \nicer and \nustar spectra, we fixed the photon‑index offsets to $d\Gamma_{1}$ = $d\Gamma_{2}$ = 0, and the break energy to $E_{br}$ = 1\,keV. This effectively reduces the correction to a simple constant factor, equivalent to the \texttt{CONSTANT} model. To establish a reliable absolute flux scale, we fixed the normalization of the \nustar FPMA module to unity ($Norm$ = 1), while leaving the normalizations of the other modules free to vary.
For the \ixpe data, we allowed all \texttt{MBPO} parameters to vary freely for each DU.
The best‑fitting \texttt{MBPO} parameters are listed in Table~\ref{tab:mbpo}. We note a $d\Gamma_1$ consistent with zero and a change in the index at ${\sim}3$\,keV in all three \ixpe DUs.

\begin{table}[!htb]
\centering
\caption{Best-fit parameters of the \texttt{MBPO} model. The values listed correspond to the joint fit described in Table~\ref{tab:Best-fit_Parameters}.
Parameters in squared brackets are fixed in the fit.
All uncertainties are reported at the 68\%\,CL.}
\label{tab:mbpo}
\begin{tabular}{l c c}
\hline
Data & Parameter (unit) & Value \\
\hline
\nicer 5100070101 & $norm$ & $0.953\pm0.005$ \\
\nicer 5100070102 & $norm$ & $0.986\pm0.005$ \\
\nicer 5100070105 & $norm$ & $1.008\pm0.005$ \\
\nicer 5100070106 & $norm$ & $0.917\pm0.005$ \\
\nicer 5100070109 & $norm$ & $0.942\pm0.005$ \\
\nicer 5100070110 & $norm$ & $0.941\pm0.005$ \\[0.5em]
\nustar FPMA & $norm$ & [1.000] \\
\nustar FPMB & $norm$ & $0.992\pm0.005$ \\[0.5em]
\ixpe DU1 & $d\Gamma_{1}$ & $0.05^{+0.11}_{-0.07}$ \\
 & $d\Gamma_{2}$ & $-0.35\pm0.03$ \\
 & $E_{br}$ (keV) & $2.6\pm0.1$ \\
 & $norm$ & $1.09^{+0.02}_{-0.01}$ \\[0.5em]
\ixpe DU2 & $d\Gamma_{1}$ & $0.05^{+0.06}_{-0.05}$ \\
 & $d\Gamma_{2}$ & $-0.37\pm0.04$ \\
 & $E_{br}$ (keV) & $2.9^{+0.2}_{-0.1}$ \\
 & $norm$ & $1.11\pm0.01$ \\[0.5em]
\ixpe DU3 & $d\Gamma_{1}$ & $0.04^{+0.04}_{-0.03}$ \\
 & $d\Gamma_{2}$ & $-0.43^{+0.05}_{-0.06}$ \\
 & $E_{br}$ (keV) & $3.2^{+0.1}_{-0.2}$ \\
 & $norm$ & $1.08\pm0.01$ \\
\hline
\end{tabular}
\end{table}

\end{appendix}

\end{document}